\PassOptionsToPackage{dvipsnames,table}{xcolor}
\documentclass[unnumsec,webpdf,modern,large]{mam-authoring-template}%

\graphicspath{{Fig/}}

\usepackage[mathlines, switch]{lineno}
\usepackage[version=3]{mhchem} %
\usepackage{booktabs}
\usepackage{amsmath,amssymb}
\usepackage{bm}
\usepackage[normalem]{ulem}
\usepackage{pdfpages}
\usepackage{mathtools}
\usepackage{placeins} %
\usepackage{amssymb}  %
\usepackage{amsmath}  %
\usepackage{threeparttable}
\usepackage{lmodern} %
\usepackage{nameref}

\usepackage{hyperref}
\hypersetup{
    colorlinks=true,
    citecolor=BrickRed, %
    linkcolor=BrickRed,
    filecolor=BrickRed,      
    urlcolor=BrickRed,
    pdftitle={Overleaf Example},
    pdfpagemode=FullScreen,
    }

\usepackage{todonotes}

\theoremstyle{thmstyleone}%
\theoremstyle{thmstyletwo}%
\theoremstyle{thmstylethree}%

\begin{document}

\journaltitle{arXiv}
\DOI{DOI HERE}
\copyrightyear{2025}
\pubyear{2014}
\access{Advance Access Publication Date: Day Month Year}
\appnotes{Original Article}

\firstpage{1}

\title[Tilt-corrected 4D-STEM with Aberrations]{Using Aberrations to Improve Dose-Efficient Tilt-corrected 4D-STEM Imaging}

\author[1$\dagger$]{Desheng Ma\ORCID{0000-0001-5237-0933}}
\author[1,2]{David A. Muller\ORCID{0000-0003-4129-0473}}
\author[1,3$\dagger$]{Steven E. Zeltmann\ORCID{0000-0003-1790-3137}}

\authormark{Ma
et al.}

\address[1]{\orgdiv{School of Applied and Engineering Physics}, \orgname{Cornell University}, \state{Ithaca, New York}, \country{USA}}  
\address[2]{\orgdiv{Kavli Institute at Cornell for Nanoscale Science}, \orgname{Cornell University}, \state{Ithaca, New York}, \country{USA}} 
\address[3]{\orgdiv{Platform for the Accelerated Realization, Analysis, and Discovery of Interface Materials}, \orgname{Cornell University}, \state{Ithaca, New York}, \country{USA}}

\corresp[$\dagger$]{Corresponding authors. \href{email:dm852@cornell.edu}{dm852@cornell.edu}, \href{email:steven.zeltmann@cornell.edu}{steven.zeltmann@cornell.edu}}

\received{Date}{0}{Year}
\revised{Date}{0}{Year}
\accepted{Date}{0}{Year}

\abstract{
Tilt-corrected imaging methods in four-dimensional scanning transmission electron microscopy (4D-STEM) have recently emerged as a new class of direct ptychography methods that are especially useful at low dose. 
The operation of tilt correction unfolds the contrast transfer functions (CTF) of the virtual bright-field images and retains coherence by correcting defocus-induced spatial shifts. 
By performing summation or subtraction of the tilt-corrected images, the real or imaginary parts of the complex phase-contrast transfer functions are recovered, producing a tilt-corrected bright field image (tcBF) or a differential phase contrast image (tcDPC). 
However, the CTF can be strongly damped by the introduction of higher-order aberrations than defocus. 
In this paper, we show how aberration-corrected bright-field imaging (acBF), which combines tcBF and tcDPC, enables continuously-nonzero contrast transfer within the information limit, even in the presence of higher-order aberrations.  At Scherzer defocus in a spherically-aberration-limited system, the resultant phase shift from the probe-forming lens acts as a phase plate, removing oscillations from the acBF CTF. 
We demonstrate acBF on both simulated and experimental data, showing it produces superior performance to tcBF or DPC methods alone, and discuss its limitations.
}
\maketitle

% \begin{linenumbers}

\section{Introduction}\label{Introduction}

Phase contrast imaging in electron microscopy uses the interference between scattered and incident waves to produce contrast.
The phase contrast signal arises from phase shifts imprinted on the electron wavefront as it passes through a potential, as described by the the quantum theory of scattering, producing constructive or destructive interference when the waves are brought together at the image plane. 
In conventional phase contrast imaging techniques, the symmetry between Friedel pairs of scattered waves cancels the phase contrast, so an additional phase shift must be introduced in order to produce constructive or destructive interference.
The maximum amount of interference contrast is obtained when the scattered waves have a $\pi/2$ phase shift with respect to the unscattered beam, which is referred to as the Zernike phase condition (\cite{zernikePhaseContrastNew1942}).
The Zernike condition describes the ideal phase shifting (\cite{dwyer_quantum_2024}), which is impossible to realize in practice; the finite width of the phase-shifted region in a real system has considerable impact on the true dose-efficiency of phase plate images (\cite{vega2025retrieval}).
Experimentally, phase plates have been produced which use Coulomb interaction with physical material (\cite{danev2014volta}) or ponderomotive interaction with an intense light wave (\cite{schwartz2019laser}), but they remain somewhat difficult to use.
In common practice, the phase shift on the scattered beams is instead produced directly by the objective lens of the microscope.
The optimum conditions for maximizing the phase contrast produced this way, which balance different orders of aberrations to approximate a phase plate, have been studied since the very early days of electron microscopy (\cite{scherzer1949theoretical}).
Producing phase contrast using lens aberrations is straightforward, but the resulting images do not have uniform transfer of all spatial frequencies and so are not directly interpretable, suffering from delocalization of high resolution features (\cite{rose1974phase}). 
These conventional imaging approaches also access only the symmetric portion of the phase contrast, as any anti-symmetric scattering is always canceled when two scattered waves are interfered; accessing this information in a conventional instrument would require blocking half of the scattered waves (\cite{glaeserProspectsExtendingResolution1979}). 
Post-processing of sequences of phase contrast images can overcome some of these limitations. 
Delocalization and non-interpretable contrast can be greatly reduced through image reconstructions of focal series (\cite{kirkland1980digital, kirkland1984improved, coene1992phase, coene1996maximum}) or tilt series of images (\cite{kirkland1995super,kirkland1997multiple}). 
These approaches are able to compensate for the aberrations of the microscope (including non-round aberrations), but proved challenging due to experimental difficulties in maintaining consistency of the imaging conditions over the acquisition time and contrast degradation from partial temporal coherence.

By the theorem of reciprocity, the analysis of image formation developed for the conventional transmission electron microscope (CTEM) is also applicable to the scanning transmission electron microscope (STEM) by time reversal symmetry and exchanging the role of the source and the detector (\cite{cowley_image_1969}).
The STEM geometry also offers the possibility of producing imaging modes that are difficult to record efficiently in the CTEM (\cite{rose_nonstandard_1976}) because of the greater flexibility in configuring the detectors than in reshaping the illumination; a notable example is differential phase contrast imaging, which produces a dose-efficient and interpretable phase image by accessing the anti-Friedel scattering while utilizing the full forward scattered beam (\cite{dekkers_differential_1974}).
STEM allows the flexibility in detector configuration to be combined with the phase shaping provided by the objective lens to produce optimal imaging conditions.
\cite{hammel1995optimum} proposed matching detector zones to the phase profile of the illumination, an approach that mimics the CTF correction methods applied in TEM mode but without requiring post-processing. 
Pixelated direct electron detectors (\cite{tate_high_2016,philipp2022very,zambonHighframeRateHighcount2023,mirCharacterisationMedipix3Detector2017}) push this flexibility to its ultimate limit by recording full intensity range of the scattered beam at every scan position, an acquisition mode referred to as four dimensional STEM (4D-STEM) (\cite{ophus_four-dimensional_2019}).
In addition to reproducing the conventional STEM imaging modes (\cite{tate_high_2016,hachtel2018sub}) and allowing for a more efficient realization of DPC (\cite{muller2014atomic,liIntegratedDifferentialPhase2022}), this detailed scattering information has also enabled a family advanced image reconstruction schemes referred to as ptychography (a recent review is given by \cite{clarkElectronPtychography2025}).
Optimizing the illumination phase shifts together with the detector configuration has been revisited in the 4D-STEM mode with matched illumination-detector interferometry (MIDI) (\cite{ophus2016efficient}).
The availability of greater computational processing capabilities for 4D-STEM, along with the realization that the 4D-STEM mode provides other practical benefits for low-dose biological imaging such as reduced chromatic blur and faster imaging via upsampling (\cite{spoth_dose-efficient_2017,yu_dose-efficient_2022,yu_dose-efficient_2024, yuDoseefficientCryoelectronMicroscopy2025,varnavides2025relaxing}), has recently motivated more detailed study of how information is encoded in a 4D-STEM dataset and how to optimally produce and extract it (\cite{ma_information_2025}).

In general, ptychographic methods proceed by considering how the diffraction pattern at each point in a 4D-STEM dataset can be converted to a point (or area) in the reconstructed image. However, rather than considering a 4D-STEM dataset as a collection of diffraction patterns, we can equivalently view the dataset as a collection of STEM images formed by each of the individual pixels on the diffraction detector. 
In this view, each image formed from a single detector pixel collected in STEM mode with a convergent beam is equivalent to a plane wave image formed in CTEM mode with tilted illumination (\cite{cowley_image_1969,rose_nonstandard_1976}). 
Because the detector has many pixels (generally the direct beam in 4D-STEM covers hundreds to thousands of pixels), the angle subtended by each pixel is very small in comparison to the illumination angle and so by reciprocity these images have a very high degree of coherence, transferring phase contrast information very efficiently. 
For the same reason, and much like the situation for the focal- and tilt-series methods in CTEM, the contrast transfer of each image is a very complicated function of both the aberrations and the aperture and so computational reconstruction schemes are needed to recover a useful image from the dataset. 

Working in this image space, a simple reconstruction approach for recovering a coherent phase contrast image from a 4D-STEM dataset is tilt-corrected bright-field (tcBF) STEM.
TcBF works well at very low dose ($<$ 1 e-/$\text{Å}^2$) and through very thick ($>$ 500 nm) but weakly-scattering (weak phase, e.g., biological) samples, and via upsampling can provide reconstructions with spatial sampling finer than the raster scan of the probe (\cite{yu_dose-efficient_2022, yu_dose-efficient_2024, nguyen_4d-stem_2016}). 
A full writeup of the tcBF method was delayed by Lena Kourkoutis' long illness and untimely passing, even as the technique was adopted and built upon in the community (\cite{varnavides_iterative_2023,kucukoglu_low-dose_2024,seiferShadowMontageConeBeam2025}).
TcBF-STEM relies on the result that, when a defocus is applied and other aberrations are negligible, to a first order approximation each single-pixel bright-field image is simply a spatially shifted copy of the underlying image (\cite{cueva2021transforming, lupini_rapid_2016, seifer2021flexible, ma2025emittance}); measuring the shifts by cross-correlation and summing the appropriately un-shifted images yields the tcBF image. 
These tilt-induced image shifts have previously been used for aberration measurement in the context of microscope auto-alignment in both TEM and by reciprocity in STEM (\cite{koster_measurement_1991, koster1992practical,lupini_rapid_2016}). 
Theoretical analysis of the contrast transfer of tcBF-STEM under a weak phase approximation was presented in \cite{yu_dose-efficient_2024}, where it was shown to have a CTEM-like transfer of information: only the symmetric scattering is recovered, with oscillations and zeroes in the transfer function, but with an information limit of twice the illumination aperture. 
Analogous to CTEM, tcBF's use of defocus to produce the phase contrast yields good transfer of low spatial frequencies but the oscillatory behavior complicates analysis at high resolution and information is lost at the zero crossings. 
\cite{ma_information_2025} extended this analysis to show that the anti-symmetric scattering is simultaneously encoded in the 4D-STEM dataset, proposing a new method referred to as aberration-corrected bright-field (acBF) which utilizes both symmetric and anti-symmetric scattering contributions to recover all information encoded in the 4D-STEM measurement under the weak phase approximation. 
Because acBF-STEM also removes the oscillations and zeros in the contrast transfer function, it yields a directly interpretable phase image free of delocalization and contrast reversals.

Here, we analyze the performance of aberration-corrected and tilt-corrected 4D-STEM imaging methods under the influence of aberrations in addition to defocus. 
We first consider the use of spherical aberration to produce a Scherzer phase condition in the illumination.
When spherical aberration is present, we recover the well-known result that the off-axial detector images have a different effective defocus and astigmatism (\cite{zemlin_coma-free_1978}).
TcBF does not account for these changes in transfer for each virtual image, so upon summation the oscillations in the CTFs largely add out of phase and the overall contrast is strongly reduced.
AcBF, however, compensates for the variation in contrast transfer of each image before summation, effectively treating the set of virtual images as a defocus series and so yields a smooth transfer function up to the diffraction limit.
This result is particularly important in the context of cryogenic imaging of biological samples as the instruments used for such experiments are rarely corrected for spherical aberration of the probe-forming optics. 
We also consider the impact of non-round aberrations such as astigmatism; these also greatly suppress tcBF contrast but offer no benefits for acBF reconstructions.
For all aberrations beyond defocus, tcBF contrast is reduced as a result of the breaking of symmetries with respect to detector position. 
By deconvolving the influence of probe aberrations (including non-round and higher-order terms), acBF serves as an ideal non-iterative ptychography method under the weak phase assumption. 
While iterative ptychography recovers information from the dataset through an entirely different algorithm than the tilt-corrected methods we discuss, our results about the efficiency of information transfer into the raw dataset are general and so offer insight into what sample information is accessible by general retrieval methods. 
Recent work on ptychography with uncorrected microscopes (\cite{nguyen2024achieving}) also motivates this study of 4D-STEM information transfer when spherical aberration is significant.

\section{Tilt-correction unfolds the complex CTF}\label{acBF}

\begin{figure*}[h]
\centering
\includegraphics[width=0.9\linewidth]{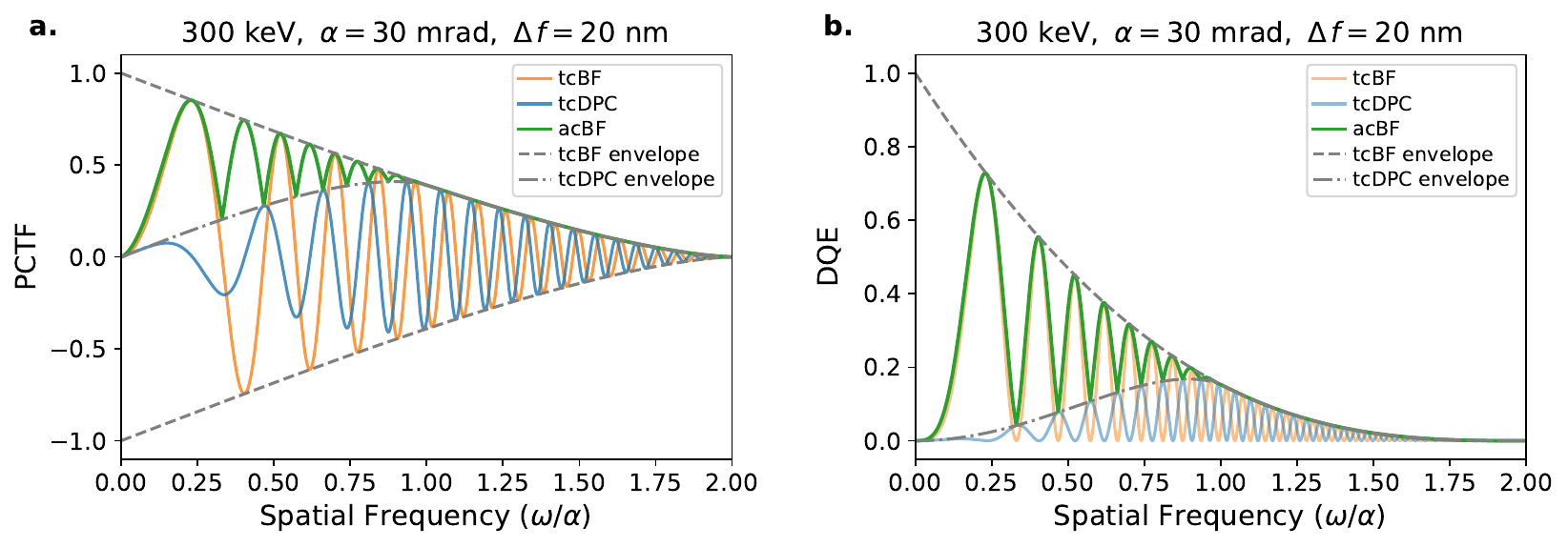}
\caption{ (a) Phase contrast transfer function (PCTF) and (b) Detective quantum efficiency (DQE)
of tcBF, tcDPC and acBF at 300 kV, 30 mrad convergence semi-angle. The tcDPC curve in (a) shows the imaginary part of its PCTF. The
tcDPC envelope is equal to the PCTF of in-focus tcDPC. acBF fills in the difference between tcBF and tcDPC, which yields constantly non-zero contrast transfer.} 
\label{1d_ctf_df}
\end{figure*}

\begin{figure*}[h]
\centering
\includegraphics[width=0.95\linewidth]{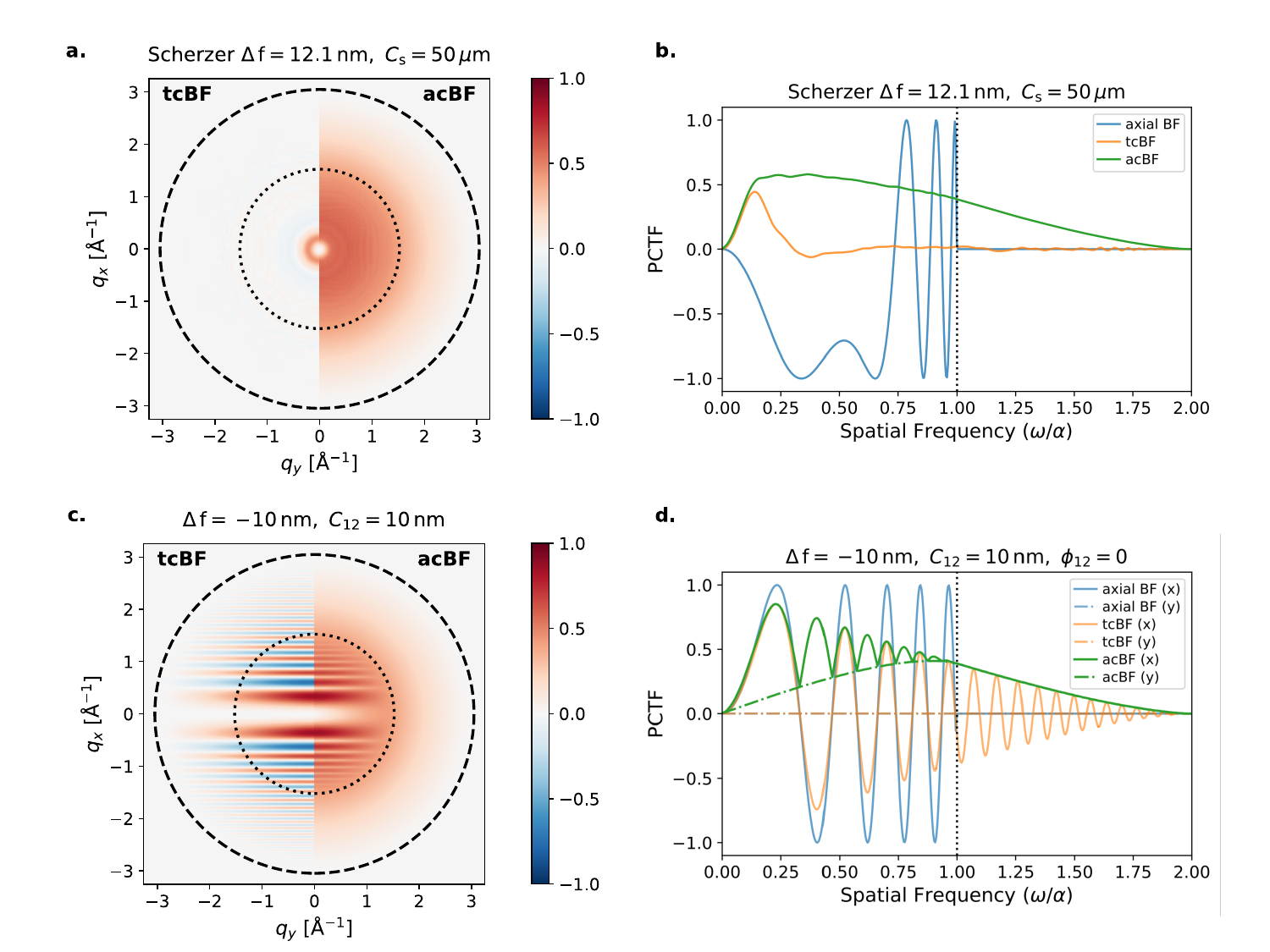}
\caption{Phase contrast transfer functions (PCTF) at 300 kV, 30 mrad convergence semi-angle in the case of (a,b) $C_s=50 \ \mu$m with Scherzer defocus and (c,d) two-fold astigmatism $C_{12}$ of 10~nm defocus of 10 nm. (b) and (d) are line profiles of the 2D PCTFs in (a) and (c), respectively. In the case of spherical aberration, the PCTF of tcBF is significantly damped due the summation over images with different effective defocus, while acBF corrects for this variation and yields a smooth PCTF. In the case of equal defocus and 2-fold astigmatism, along one direction the defocus effectively doubles, while along the other direction the defocus is zero. tcBF has no contrast along the zero-defocus axis, while acBF recovers the antisymmetric scattering only. Astigmatism of the probe leads to irrecoverable information loss in the dataset. }\label{1d_2d_ctf_sch_stig}
\end{figure*}

The general form of contrast transfer in CTEM at tilted illumination (or by reciprocity, STEM) is given by the quantum theory of electron scattering, and can include both phase and amplitude contrast. 
When the sample is a weak scatterer and thus does not significantly alter the amplitude of the incident electron beam, but only induces a small phase shift to the incident electron beam, i.e., the weak phase object approximation (WPOA), a linear phase contrast transfer function (PCTF) can be derived (\cite{kirkland_advanced_2020}).

By reciprocity, the complex phase contrast transfer function a virtual BF image formed from a single small pixel on the diffraction detector is (\cite{rose_nonstandard_1976}),
\begin{equation}\label{tilt_pctf}
\begin{aligned}
    \operatorname{PCTF}(\boldsymbol{\omega}, \boldsymbol{\Theta})=\frac{i}{2 \Omega_0} A(\boldsymbol{\Theta})\bigl\{A(\boldsymbol{\omega}-\boldsymbol{\Theta}) e^{-i[\chi(\boldsymbol{\omega}-\boldsymbol{\Theta})-\chi(\boldsymbol{\Theta})]}\\-A(\boldsymbol{\omega}+\boldsymbol{\Theta}) e^{+i[\chi(\boldsymbol{\omega}+\boldsymbol{\Theta})-\chi(\boldsymbol{\Theta})]}\bigr\},
\end{aligned}
\end{equation}
where $\boldsymbol{\omega}$ is the spatial frequency, i.e., scattering vector, $\Omega_0$ is the solid angle of the illumination cone, and $\boldsymbol{\Theta}$ is the position (i.e., pixel) on the detector. $A(\boldsymbol{\Theta})$ is the aperture function equal to 1 inside the aperture $|\boldsymbol{\Theta}| < \alpha$ and 0 outside. $\chi$ is the phase shift of the scattered electrons attributable to the lens aberrations and is given in polar coordinates by
\begin{equation}
\begin{split}
    \chi(\theta, \phi)&=\frac{2 \pi}{\lambda} \sum_{n, m} \frac{C_{nm} \theta^{n+1} \cos \left(m\left(\phi-\phi_{nm}\right)\right)}{n+1}\\
    &\stackrel{or}{=}\frac{2 \pi}{\lambda} \sum_{n,m} \frac{\theta^{n+1}}{n+1}\left|\tilde{C}_{nm}^{*} e^{im\left(\phi-\phi_{n m}\right)}\right|,
\end{split}
\label{equ: abr}
\end{equation}
where the coefficients $\tilde{C}_{nm} = C_{nm} e^{{m \phi_{nm}}}$ refer to aberrations with angular order $n$ and $m$-fold azimuthal symmetry, 
and $\phi_{nm}$ is the azimuth for the non-round aberrations.

\subsubsection{Tilt-corrected imaging with defocus only}
First, we will briefly review the existing results for the case when defocus is the only aberration present, for which analytical CTFs can be derived for the final tilt-corrected images. 
Keeping only $C_{10} = -\Delta f$, the aberration function becomes $\chi(\bm\omega) = \frac12 k_0 \Delta f |\bm\omega|^2$ and thus Equation~\ref{tilt_pctf} reduces to

\begin{equation}
\label{tcBF_df_shift}
\resizebox{\columnwidth}{!}{$
\begin{gathered}
\operatorname{PCTF}^{(\Delta f)}(\boldsymbol{\omega}, \boldsymbol{\Theta})
= \frac{i}{2\Omega_{0}}\,A(\boldsymbol{\Theta}) \Bigl\{ 
    A(\boldsymbol{\omega}-\boldsymbol{\Theta}) 
    \exp\!\left( \tfrac{i}{2} k_{0}\,\Delta f\, |\boldsymbol{\omega}|^{2} \right) \\
\quad\;\; - 
    A(\boldsymbol{\omega}+\boldsymbol{\Theta}) 
    \exp\!\left( -\tfrac{i}{2} k_{0}\,\Delta f\, |\boldsymbol{\omega}|^{2} \right) 
\Bigr\} 
\exp\!\left[ -i \bigl(\Delta f\,\boldsymbol{\Theta}\bigr)\cdot \left(k_{0}\boldsymbol{\omega}\right) \right],
\end{gathered}
$}
\end{equation}
where $k_0 = \frac{2\pi}{\lambda}$ is the wave number of the incident electrons.\\
The factor $\exp\!\left[ -i \bigl(\Delta f\,\boldsymbol{\Theta}\bigr)\cdot \left(k_{0}\boldsymbol{\omega}\right) \right] $ in this PCTF, by the Fourier shift theorem, corresponds to a real space image shift equal to $\Delta f \mathbf{\Theta}$.
The PCTF here is conjugate symmetric with respect to $\mathbf{\Theta}$,
\begin{equation}\label{conjugate}
\operatorname{PCTF}^{(\Delta f)}(\boldsymbol{\omega},-\boldsymbol{\Theta})^*=\operatorname{PCTF}^{{(\Delta f)}}(\boldsymbol{\omega},+\boldsymbol{\Theta}).
\end{equation}

Summation over all shift-corrected images within the bright-field disk yields tcBF, whose PCTF is derived in \cite{yu_dose-efficient_2024}:

\begin{equation}
    \begin{aligned}
\operatorname{PCTF}_{\mathrm{tcBF}}^{(\Delta f)}({\boldsymbol{\omega}}) 
& =-\mathbb{A}(\boldsymbol{\omega}) \sin \left(\frac{1}{2} k_0 \Delta f |\boldsymbol{\omega}|^{2}\right).
\end{aligned}
\end{equation}

Here $\mathbb{A}(\boldsymbol{\omega})$ is the normalized area of the overlap region between two disks separated by spatial frequency $|\boldsymbol{\omega}| \in[0,2 \alpha]$, or can be equivalently viewed as an aperture correlation function (\cite{yang2015efficient, ma_information_2025}.) TcBF thus can have an information limit up to $|2 \alpha|$, doubling that of axial BF-STEM. Similar to axial BF, tcBF requires defocus to generate contrast. An example of this contrast transfer function is shown in Figure 1a.

The conjugate symmetry (Eq.~\ref{conjugate}) implies that summation of Friedel pairs of images eliminates any imaginary component in the PCTF.
This component, arising from the anti-Friedel scattering, can be extracted by instead subtracting Friedel pairs of shift-corrected images to produce a tilt-corrected differential phase contrast image, i.e., tcDPC (\cite{ma_information_2025}), which has the transfer function
\begin{equation}
\begin{aligned}
\operatorname{PCTF}_{\mathrm{tcDPC}}^{(\Delta f)}(\boldsymbol{\omega}) 
& =i[\mathbb{A}(\boldsymbol{\omega})-\mathbb{A}(2 \boldsymbol{\omega})] \cos \left(\frac{1}{2} k_0 \Delta f |\boldsymbol{\omega}|^{2}\right).
\end{aligned}
\end{equation}

The PCTF of tcDPC is purely imaginary, producing a gradient or differential phase image in real space, as in conventional DPC, and is maximized at zero defocus. 
Without consideration of noise, in-focus tcDPC yields the same information transfer as in-focus single sideband ptychography (SSB) (\cite{pennycook_efficient_2015, yang2015efficient}) as identified by \cite{ma_information_2025}.

The $\sin \left(\frac{1}{2} k_0 \Delta f |\boldsymbol{\omega}|^{2}\right)$ modulation of tcBF and the $\cos \left(\frac{1}{2} k_0 \Delta f |\boldsymbol{\omega}|^{2}\right)$ modulation of tcDPC (along with their different envelope functions) are exactly orthogonal to each other and suggest that an optimally efficient imaging method can be produced by appropriately combining the images formed from each method, especially since zero crossings
in tcBF will be where tcDPC is maximal and vice versa. The combination effectively allows for aberration (or CTF) correction of each tilted-angle image before summation, which we term aberration-corrected bright-field imaging (acBF). 
The analytical PCTF for acBF in the case of only defocus is (\cite{ma_information_2025}):
\begin{equation}\label{acbf_pctf_analytical}
\operatorname{PCTF}_{\mathrm{acBF}}^{(\Delta f)}(\boldsymbol{\omega})=[\mathbb{A}(\boldsymbol{\omega})-\mathbb{A}(2 \boldsymbol{\omega})]+\mathbb{A}(2 \boldsymbol{\omega})\left|\sin \left( \frac{1}{2}k_0 \Delta f |\boldsymbol{\omega}|^{2} \right)\right|.
\end{equation}

It is clear from the second term of Equation \ref{acbf_pctf_analytical} that defocus produces extra contrast transfer in addition to in-focus tcDPC by using the symmetric scattering components. 
A comparison of the CTFs of the three tilt-corrected imaging modes is presented in Figure \ref{1d_ctf_df}a.

The CTF describes information transfer at different spatial frequencies under an infinite signal-to-noise ratio (SNR) assumption. Evaluation of dose efficiency needs to incorporate noise, which is described by the detective quantum efficiency (DQE) (\cite{mcmullan_detective_2009}):

\begin{equation}
\operatorname{DQE}(\boldsymbol{\omega})=\operatorname{DQE}(0) \frac{|\operatorname{PCTF}(\boldsymbol{\omega})|^2}{|\operatorname{NPS}(\boldsymbol{\omega})|^2},
\end{equation}
where $\operatorname{DQE}(0)=1$ in a perfect pixel detector and $\operatorname{NPS}(\boldsymbol{\omega})$ is the noise power spectrum. For an ideal detector the measurement noise is only from Poisson noise. For bright-field imaging of a weakly scattering object, the NPS can be approximated as flat. More discussions on the DQE can be found in \cite{ma_information_2025} and \cite{bennemannDetectiveQuantumEfficiency2025}, including the propagation of noise through more complicated reconstruction techniques. A comparison of the DQEs of the three tilt-corrected imaging modes is presented in Figure \ref{1d_ctf_df}b.

\subsubsection{Aberrations beyond defocus}
Tilt-corrected imaging is not limited to the defocus-only case.
For other aberrations, the image shifts have been previously derived by taking a Taylor expansion of Equation~\ref{tilt_pctf} with respect to $\chi$ or by taking the Wigner-Weyl transform of the electron wavefunction \cite{ma2025emittance}.
Both these approaches give the image shift $\Delta \rho$ of an off-axial STEM image as (\cite{lupini_rapid_2016})
\begin{equation}
    \Delta \rho(\bm{\Theta})= \nabla \chi(\bm\Theta)
    \label{eq:lupini-shift}
\end{equation}
where $\nabla$ is the gradient operator. 
However, this approximation does not yield insight into the contrast transfer of the summed images. 
The conjugate symmetry relation \ref{conjugate} does not hold for all aberrations, so it is not generally the case that the tcBF PCTF will be the corresponding
CTEM PCTF multiplied by the envelope function $\mathbb{A}(\boldsymbol{\omega})$. 
Additional cross-terms in $\chi(\bm{\omega}\pm\bm{\Theta})-\chi(\bm{\Theta})$ appear for aberrations other than defocus, which add a complicated tilt- and azimuth-dependence to the contrast transfer of the off-axial images. 

To illustrate this breakdown, we consider two representative cases of including higher order aberrations than defocus, (i) spherical aberration $C_s \equiv C_{30} $ and (ii) 2-fold astigmatism $C_{12}$. 
We also derive the results for the second order aberrations (three-fold astigmatism $A_2 \equiv C_{23}$, axial coma $B_2 \equiv C_{21}$) in \hyperref[appendix_A]{Appendix~A}.

\subsubsection{Spherical aberration}

In the case of (i) $C_{30}$, the PCTF for the image from a single detector pixel is given by

\begin{equation}
\label{C3_pctf}
\resizebox{\columnwidth}{!}{$
\begin{gathered}
\mathrm{PCTF}^{(C_{30})}(\boldsymbol{\omega},\boldsymbol{\Theta})
=\frac{i}{2\Omega_0}\,A(\boldsymbol{\Theta})\;
\exp\!\Bigg\{\,i k_0 C_{30}\,(|\boldsymbol{\omega}|^2+|\boldsymbol{\Theta}|^2)\left(\boldsymbol{\omega}\cdot \boldsymbol{\Theta}\right)\Bigg\} \\[6pt]
\quad\times\Bigg[
A(\boldsymbol{\omega}-\boldsymbol{\Theta})\,
\exp\!\Bigg(-ik_0\frac{C_{30}}{4}\,
\Big[|\boldsymbol{\omega}|^4+4\left(\boldsymbol{\omega}\cdot\boldsymbol{\Theta}\right)^2+2|\boldsymbol{\omega}|^2|\boldsymbol{\Theta}|^2\Big]\Bigg) \\[6pt]
\qquad-\;
A(\boldsymbol{\omega}+\boldsymbol{\Theta})\,
\exp\!\Bigg(+ik_0\frac{C_{30}}{4}\,
\Big[|\boldsymbol{\omega}|^4+4\left(\boldsymbol{\omega}\cdot\boldsymbol{\Theta}\right)^2+2|\boldsymbol{\omega}|^2|\boldsymbol{\Theta}|^2\Big]\Bigg)
\Bigg].
\end{gathered}
$}
\end{equation}

In agreement with the linear expansion (Equation~\ref{eq:lupini-shift}), a shift of $(C_{30} |\bm{\Theta}|^2)$ and in the direction of $\bm\Theta$ factors out of the PCTF.
However, even after correction of the shifts there are additional alterations in the PCTF. 
To clarify the meaning of the additional terms, it is illuminating to evaluate Equation \ref{C3_pctf} for the triple overlap region (where the aperture terms $A(\bm\omega)$ are all 1).
Including both $C_{10}$ and $C_{30}$, the PCTF in the triple overlap is

\begin{equation}\label{acBF_sch_df_to}
\resizebox{\columnwidth}{!}{$
\begin{gathered}
\mathrm{PCTF}_{\text{TO}}^{(C_{10}+C_{30})}(\boldsymbol{\omega},\boldsymbol{\Theta})
= \frac{1}{\Omega_0}
\underbrace{e^{ ik_0\left(C_{10}+C_{30} |\boldsymbol{\Theta}|^2\right)
\left(\boldsymbol{\omega}\cdot\boldsymbol{\Theta}\right)}}_{\text{shift}}
\underbrace{e^{i{k_0} { C_{30}}|\boldsymbol{\omega}|^2
\left(\boldsymbol{\omega}\cdot\boldsymbol{\Theta}\right)}}_{\text{distortion}} \\\times 
\sin \bigg(\tfrac{k_0}{2} \bigg[
\underbrace{\left(C_{10}+2C_{30}|\boldsymbol{\Theta}|^2\right) |\boldsymbol{\omega}|^2}_{\text{defocus}}
+\underbrace{C_{30} |\boldsymbol{\Theta}|^2 |\boldsymbol{\omega}|^2\cos 2\Delta}_{\text{astigmatism}}\bigg]
+\underbrace{\tfrac{k_0}{4}C_{30}|\boldsymbol{\omega}|^4}_{\text{spherical}}
\bigg),
\end{gathered}
$}
\end{equation}
where $\Delta = \angle(\bm\omega)-\angle(\bm\Theta)$ is the angle between the scattering vector and the detector position.

Shift correction removes the term labeled ``shift''. It is clear that the shift-corrected PCTF is very different from $\sin (\chi)$. 
The ``distortion'' term, $\exp(i k_0 C_{30} |\bm\omega|^2 (\bm\omega\cdot\bm\Theta))$, yields a complicated point spread function; due to the quadratic frequency dependence, \cite{koster1992practical} refer to this as a ``dispersive'' image shift.
While this has a complicated impact on the single pixel image contrast, we note that this term only imparts a frequency-dependent phase factor and so can be fully corrected without loss of contrast transfer efficiency so long as the spherical aberration coefficient can be estimated to sufficient accuracy.
This term does not depend on the defocus, and becomes more severe towards the edge of the aperture. 
The argument to the $\sin$ function, which can be seen as the ``effective'' aberration function for the off-axial image, is also modified by the cross-terms between $C_{10}$ and $C_{30}$. 
First, the effective defocus has increased by an amount $2C_{30}|\bm\Theta|^2$.
Additionally, a first-order astigmatism has appeared, with magnitude $C_{30}|\bm\Theta|^2$ and in the direction of $\bm\Theta$. 
This term arises despite the original aberration function having round symmetry, and is an interaction between the different radial orders. 
Finally, the spherical aberration term in the original aberration function is simply transferred to the effective aberration function unchanged. 
The appearance of the additional defocus and astigmatism have been studied before in the context of CTEM imaging with a tilted beam (\cite{zemlin_coma-free_1978,zemlin1978image}), but most classical papers on this subject perform calculations in units of defocus normalized by $C_{30}$; this scaling was common in the era before aberration correction but cannot be applied when $C_{30} \rightarrow 0$. 
Because these modifications are arguments to the $\sin$ function, these terms are able to damp (or enhance) the overall magnitude of the contrast transfer and so have an impact that cannot be completely removed by tcBF or acBF. 
This situation applies to the triple overlap region, and so impacts the transfer for frequencies $|\bm\omega|<\alpha$ (though note that the double overlap also contributes in this region). 
In the double overlap regions, where only one of the shifted aperture terms is nonzero, Equation~\ref{C3_pctf} maintains unit magnitude (but retains its complicated phase modulation); full recovery of information remains possible in the DO region.
This is illustrated in \hyperref[appendix_B]{Appendix~B}, where we plot the 2D $\mathrm{PCTF}(\bm\omega)$ for several $\bm\Theta$. 
The magnitude of the PCTF in the TO region oscillates from $-1$ to $ 1$ while it is uniformly $\frac12$ within the DO.
Because these terms also break the conjugate symmetry with respect to $\bm\Theta$, they also make it challenging to obtain an analytic expression for the final tcBF or acBF images.

Figure~\ref{1d_2d_ctf_sch_stig}a and b illustrate an example of the tcBF and acBF PCTFs for nonzero $C_{30}$ and Scherzer defocus $C_{10} = \sqrt{\frac43 C_{30} \lambda}$.

\subsubsection{Two-fold astigmatism}

In the case of (ii) $C_{12}$, the PCTF is given by,

\begin{equation}\label{acBF_C12}
\resizebox{\columnwidth}{!}{$
\begin{gathered}
\mathrm{PCTF}^{(C_{12})}(\boldsymbol{\omega},\boldsymbol{\Theta})
=\frac{i}{2\Omega_0}\,A(\boldsymbol{\Theta})\;
\exp\!\Bigg\{\, i k_0
\Big(\underbrace{C_{12}|\boldsymbol{\omega}||\boldsymbol{\Theta}|\,
\cos(\phi+\Phi-2\phi_{12})}_{\text{shift}}\Big)\Bigg\}
\\[4pt]
\quad\times\Bigg[
A(\boldsymbol{\omega}-\boldsymbol{\Theta})\,
\exp\!\Bigg(\!-\, i\,\frac{k_0}{2}\;
\underbrace{C_{12}|\boldsymbol{\omega}|^{2}
\cos\!\big(2(\phi-\phi_{12})\big)}_{\text{2-fold astigmatism}}
\Bigg)
\\[2pt]
\qquad-\;
A(\boldsymbol{\omega}+\boldsymbol{\Theta})\,
\exp\!\Bigg(\!+\, i\,\frac{k_0}{2}\;
\underbrace{C_{12}|\boldsymbol{\omega}|^{2}
\cos\!\big(2(\phi-\phi_{12})\big)}_{\text{2-fold astigmatism}}
\Bigg)
\Bigg],
\end{gathered}
$}
\end{equation}
where $\phi = \angle(\bm\omega)$, $\Phi=\angle(\bm\Theta)$ and 
$
\phi_{12}$ is the azimuth of the two-fold astigmatism (Eq.~\ref{equ: abr}).

Note that the astigmatism terms only depend on spatial frequency and the aberration coefficient, and do not contain any dependence on the detector position $\bm\Theta$.
This is also visible in the tableau of PCTFs in \hyperref[appendix_B]{Appendix~B}.
As above, it is illuminating to evaluate Equation \ref{acBF_C12} for the triple overlap region (where the aperture terms $A(\omega)$ are all 1).
Including both $C_{10}$ and $C_{12}$, the PCTF in the triple overlap is

\begin{equation}\label{acBF_C12_to}
\resizebox{\columnwidth}{!}{$
\begin{aligned}
\mathrm{PCTF}^{(C_{10}+C_{12})}_{\mathrm{TO}}(\boldsymbol{\omega},\boldsymbol{\Theta})
&= \frac{1}{\Omega_0}\,
\underbrace{\exp\!\Big(
i\,k_0\big[
C_{10}(\boldsymbol{\omega}\!\cdot\!\boldsymbol{\Theta})
+ C_{12}|\boldsymbol{\omega}||\boldsymbol{\Theta}|\,
\cos(\phi+\Phi-2\phi_{12})
\big]
\Big)}_{\text{shift}} \\[6pt]
&\quad \sin\!\Big(
\frac{k_0}{2}\big[
\underbrace{C_{10}|\boldsymbol{\omega}|^{2}}_{\text{defocus}}
+ \underbrace{C_{12}|\boldsymbol{\omega}|^{2}\cos\!\big(2(\phi-\phi_{12})\big)}_{\text{astigmatism}}
\big]
\Big).
\end{aligned}
$}
\end{equation}

As expected, the image shift depends on the local gradient of the lens aberration function.
The effective aberration function (argument of the $\sin$ function) notably does not depend on the detector pixel chosen, and unlike in the previous case there are no cross-terms between the two aberrations. 
The astigmatism term $\frac{1}{2}
C_{12}|\boldsymbol{\omega}|^{2}\,
{\cos\!\big(2(\phi-\phi_{12})\big)}$ is effectively a direction-dependent defocus. 
Thus, along one direction defocus and $C_{12}$ add together and along the orthogonal direction they subtract. 
Therefore, the transfer of low-frequency will vary azimuthally due to astigmatism and the lost information is not recoverable by either tcBF or acBF. 
In the DO region, the astigmatism produces a phase shift but does not damp the magnitude of information transfer and so can potentially be recovered. 
Detailed derivations of the two above cases, as well as the results for axial coma ($C_{21}$) and 3-fold astigmatism ($C_{23}$), are presented in \hyperref[appendix_A]{Appendix A}.

Figure~\ref{1d_2d_ctf_sch_stig}c and d illustrate an example of the tcBF and acBF PCTFs for the case where $C_{10} =C_{12}=-\Delta f $.
When both aberrations are of equal magnitude, the effective defocus is doubled along one direction and zero along the orthogonal direction.
As a result, tcBF produces no contrast along the $q_x=0$ line, because no TO signal is produced at zero defocus and tcBF does not recover the anti-Friedel scattering in the DO.
acBF shows good low-frequency transfer along the direction with large effective defocus and poor (DPC-like, with no TO contribution) but non-zero transfer in the other direction.

\begin{figure*}[h]%
\centering
\includegraphics[width=0.82\linewidth]{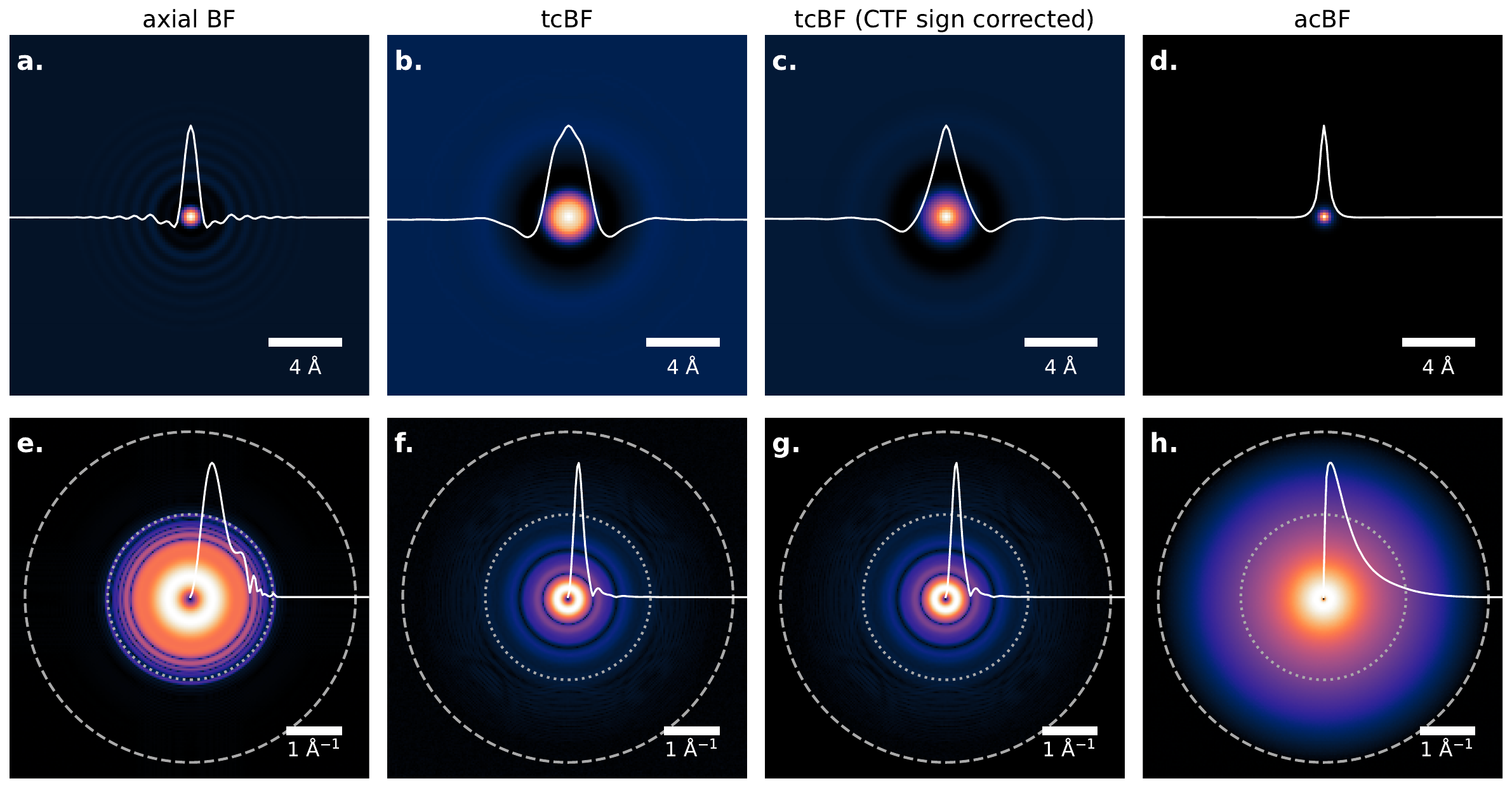}
\caption{Imaging of a simulated single C atom at infinite dose, 300 keV and 30 mrad convergence semi-angle
with $C_s = 50\ \mu$m and Scherzer defocus (121 Å) using different methods: (a) axial BF, (b) tcBF, (c) tcBF with post-summation CTF correction
(sign flipping), (d) acBF. (e–h) Corresponding fast Fourier transform (FFT) power spectra of (a–d). The dashed circles on the FFTs indicate the spatial frequencies corresponding to 1$\alpha$ and 2$\alpha$, and line traces through the center of each plot are overlaid. The tcBF images both show worse resolution and more substantial tails than the axial image due to the summation over images with different defocus and astigmatism. acBF has a sharp response and no negative tails which would cause contrast delocalization, and smooth transfer of information at all frequencies up to 2$\alpha$.}\label{fig_singleC}
\end{figure*}

\section{Results and Discussion}\label{Results and Discussion}

\begin{figure*}[h]%
\centering
\includegraphics[width=0.9\linewidth]{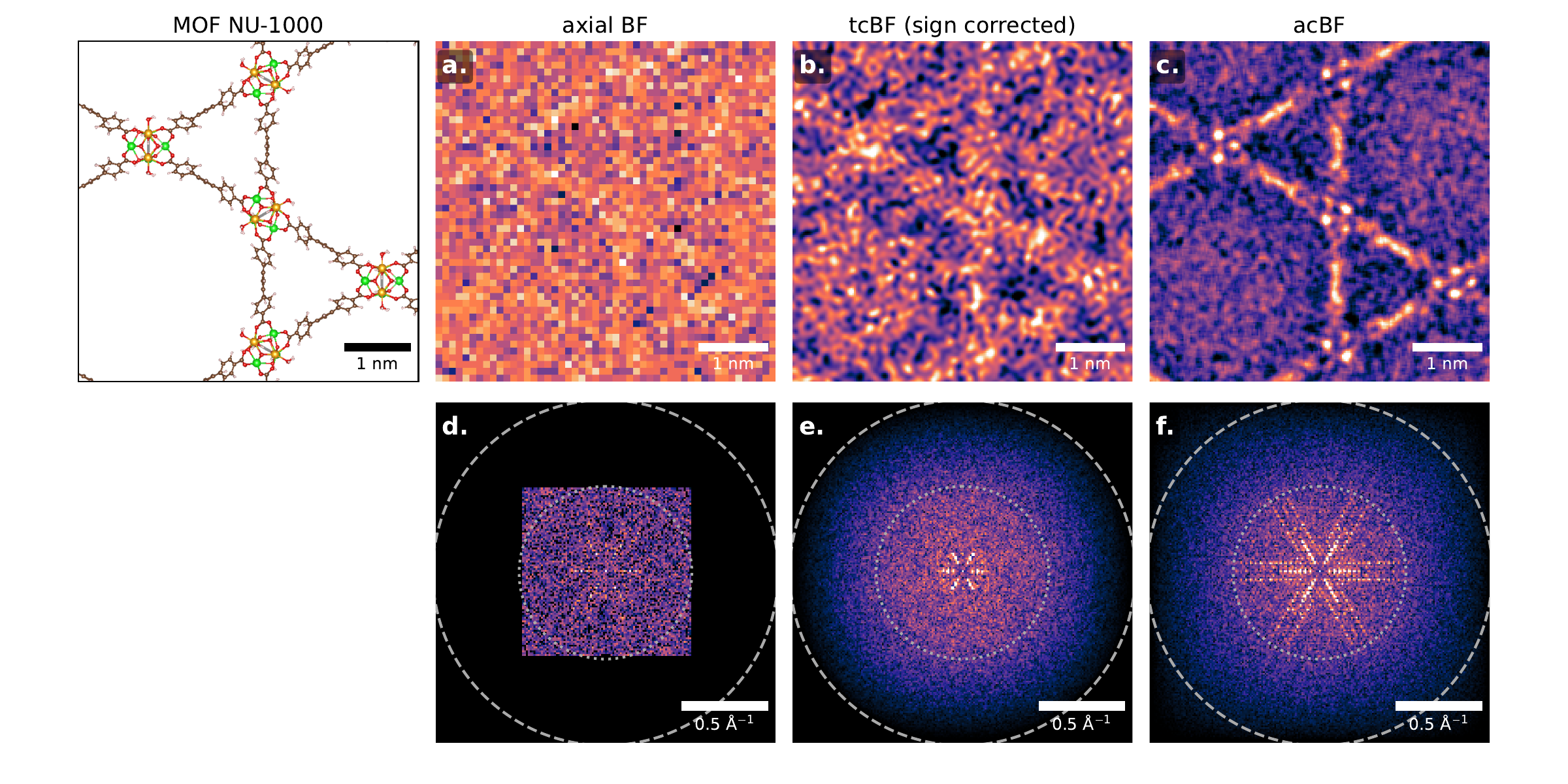}
\caption{Imaging of a simulated 3-nm-thick MOF NU-1000 sample at 200 e-/$\text{Å}^2$, 300 kV, 10 mrad semiconvergence angle, 1~\AA\ scan step size, $C_{s}$ = 2.7~mm, and  Scherzer defocus (892 \AA) to show the impact of spherical aberration using (a) axial BF, (b) tcBF with post-summation CTF correction (sign flipping), (c) acBF. TcBF and acBF have been upsampled by a factor of 4. (d–f) Corresponding fast Fourier transform (FFT) power spectra of (a–c). The dashed circles on the FFTs indicate the spatial frequencies corresponding to 1$\alpha$ and 2$\alpha$. The atomic structure schematic is displayed on the left with color denoting Zr (gold/green), O (red), C (brown) and H (pink). Within each Zr cluster, the Zr pairs with strong and weak projected electrostatic potential intensity are labeled in gold and green, respectively. The structure is barely resolvable with tcBF, while acBF resolves the structure clearly and locates the Zr atom pairs in the nodes.}\label{fig_simulated_MOF}
\end{figure*}

\begin{figure*}[!t]%
\centering
\includegraphics[width=0.822\linewidth]{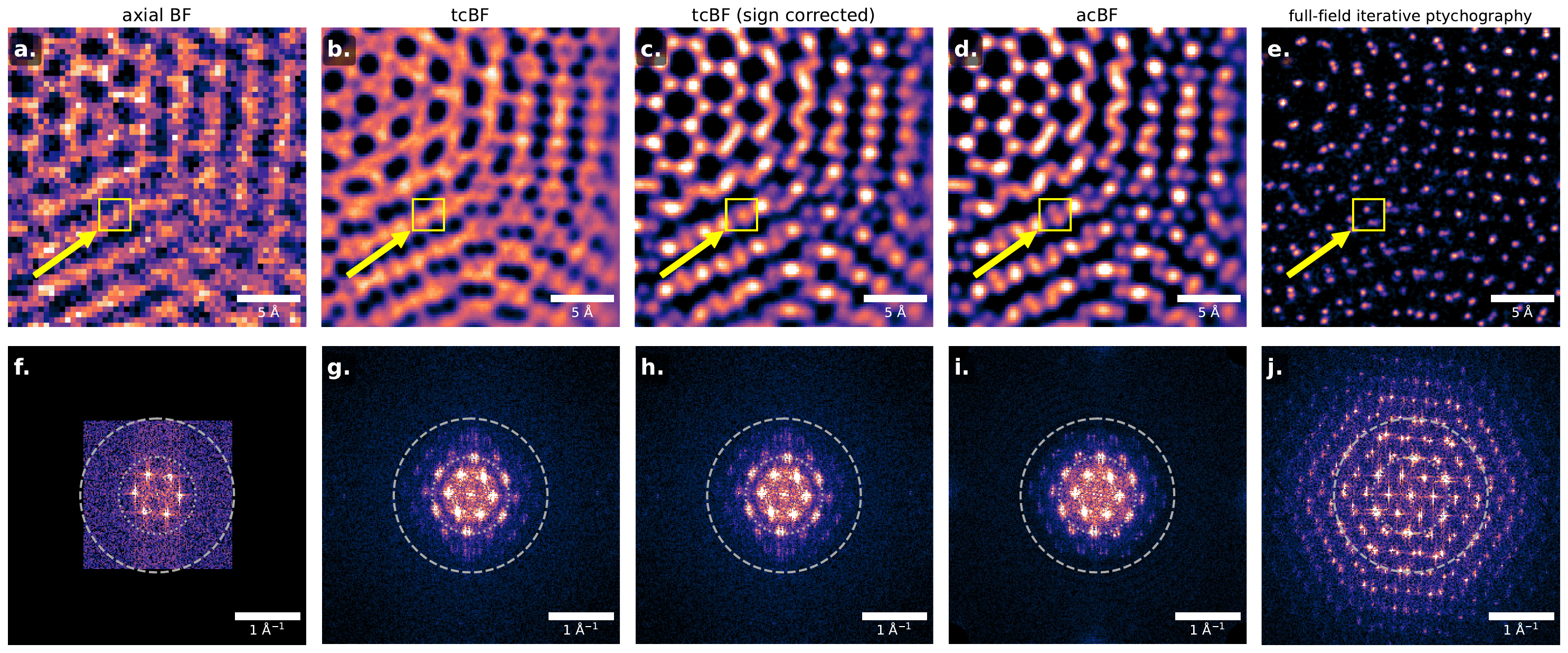}
\caption{
Reconstructed images of an experimental 4D-STEM dataset (adapted from \cite{zhang2025atom}) of a twisted tungsten diselenide (WSe$_2$) bilayer acquired at 80 kV and 25 mrad semi-convergence angle with an aberration-corrected instrument, using:
(a) axial BF, (b) tcBF, (c) tcBF with post-summation CTF correction
(sign flipping), (d) acBF, along with their respective Fourier
transforms (f-h). 
TcBF and acBF have been upsampled by a factor of 2. 
The iterative, multislice ptychography reconstruction performed by \cite{zhang2025atom} is shown in (e) with its FFT in (j) for comparison to indicate the true atomic positions. 
The yellow arrow highlights a region where two W atoms closely spaced in projection are resolved by acBF but not tcBF. 
The dashed circles on the FFTs indicate the spatial
frequencies corresponding to 1$\alpha$ and 2$\alpha$, and the tilt corrected images show an information limit of better than 1~\AA.}
\label{fig_phason}
\end{figure*}

To demonstrate the information transfer of the various tilt-corrected imaging methods, we perform a simulation study on a single C atom at infinite dose; a simulation study on a metal organic framework (MOF) sample at low dose; and reconstruction of experimental data from a twisted bilayer of tungsten diselenide ($\text{WSe}_\text{2}$) (dataset from \cite{zhang2025atom}). These examples demonstrate the point spread function, the contrast transfer at low frequencies ($|\bm\omega|<\alpha$), and the contrast transfer at high frequencies ($\alpha$--$2 \alpha$) respectively. In addition to tcBF and acBF, we also show results from ``sign-corrected'' tcBF, as suggested by \cite{varnavides_iterative_2023}, which simply inverts the contrast at spatial frequencies where the PCTF for axial imaging is negative as commonly used in cryo-EM (\cite{downing2008restoration}). 

\subsubsection{Single atom simulations}
Figure \ref{fig_singleC} shows tilt-corrected images of a single carbon atom in free space, using a spherical aberration of 50~$\mu$m and Scherzer focus with a 300 kV accelerating voltage and 30~mrad semiconvergence angle.
While this magnitude of spherical aberration is smaller compared to typical values in an uncorrected microscope (generally a few mm), we note that it is over 40~times greater than the tolerance for minimum probe size for ADF-STEM imaging with this aperture size, $C_{30}^{(\text{tol})} = \lambda / 2 \alpha^4 \approx 1.2 \, \mu\text{m}$ (\cite{kirkland_advanced_2020} Eq. 2.28). 
The single atom is effectively a pure weak phase object, and nearly infinitely small at the scale of the simulation so that the resulting images are effectively the point spread function (PSF) of the imaging system. 
The axial BF image, produced from only the central pixel on the detector, shows a large flat passband in its FFT as a result of using Scherzer focus.
Several oscillations of the CTF are included within the aperture, leading to corruption of the high spatial frequencies.
The tcBF image shows substantially worse resolution and neagative tails at a further distance from the atom as compared to the axial BF image.
This is a consequence of the variation in effective defocus and astigmatism for each of the virtual BF images from the different detector pixels not being compensated during summation, and shows that tilt correction does not always result in an improved image.
In a real experiment at very low dose the tcBF image may still be an improvement due to its superior collection efficiency compared to axial BF-STEM, but under these conditions the CTEM image with identical aperture would be superior to tcBF as it would have the same CTF as the axial BF-STEM image but full collection efficiency.
Sign correction of the tcBF CTF can moderately improve the sharpness, but it does not reduce the tails or improve the information limit. 

In contrast, acBF is unaffected and produces continuously nonzero contrast transfer up to $2\alpha$ by accounting for the variation in the PCTF of each individual tilted beam image before summation, correcting for both the dispersive shifts and the focus and astigmatism variation.
In fact, the presence of spherical aberration has smoothed the transfer function at low frequencies in $\big[0,\alpha\big]$, agreeing with the theoretical PCTFs shown in Figure \ref{1d_2d_ctf_sch_stig}c.
In the case of Scherzer focus, for a wide range of the aperture the aberration function is designed to be as flat as possible, which means that the shifts are nearly constant according to Equation~\ref{eq:lupini-shift}.
However, the effective defocus and astigmatism of the shifted images varies even within the range where the shifts are small; this effectively produces an in-line focal-series dataset in addition to the tilt-series common to all of the tilt-corrected 4D-STEM methods. 
This makes acBF with spherical aberration akin to the classical image reconstruction methods for CTEM (\cite{kirkland1980digital,kirkland1988high}) but with hundreds of combinations of tilt and defocus as compared to few to tens of images for the conventional methods; this helps overcome the limitations of the conventional methods, particularly with regards to the robustness of the esimation of the aberration function.
The images are also acquired simultaneously and generally faster than in CTEM for improved rejection of environmental and instrument instability. 
The improvement in information retrieval for acBF when spherical aberration is present also appears to hold for iterative ptychography as well.
\cite{nguyen2024achieving} performed Bayesian optimization to determine the ideal aberrations for ptychographic information transfer, and found that the best result was produced with a combination of defocus and spherical aberration.
Their optimization assumed that the spherical aberration could be varied arbitrarily, and their best result was obtained with a small, positive $C_s$ of about 100~$\mu$m for 200~kV operation and a defocus of about 4 times the Scherzer condition.
Operation at defocus beyond Scherzer may be beneficial when the finite detector pixel size is taken into consideration, or the benefits may be due to nonlinear effects in iterative ptychography. 

It has generally been assumed that tcBF imaging could correct for aberrations present in the probe, so long as the shifts could be measured accurately enough.
The analysis we present here shows that this is not the case, and that tcBF has poor information transfer when any aberration beyond defocus is present as low-order non-round aberrations reduce the total information content of the 4D-STEM dataset and high-order aberrations yield cross-terms that produce distortions and contrast changes. 
Correction of the cross-term contrast changes before summation is not only required for effective reconstruction, but can actually yield improvements in the contrast transfer when the aberrations are carefully balanced. 

\subsubsection{Simulated low-dose imaging of a MOF}
Figure \ref{fig_simulated_MOF} presents tilt-corrected imaging of a simulated dataset of a metal-organic framework (MOF) at low dose and using microscope parameters typical for an uncorrected cryogenic electron microscope.
Here, we used a spherical aberration of $C_s=2.7$~mm at an accelerating voltage of 300 kV, Scherzer focus of $\Delta f = 892$~\AA, a semiconvergence angle of 10~mrad, and applied Poisson noise to model a total dose of 200~e$^-$/\AA$^2$.
This aperture size is slightly larger than the optimal Scherzer aperture of 8~mrad for these conditions (\cite{kirkland_advanced_2020}).  
The scan step size is 1~\AA, which is coarser than the Nyquist rate for $2\alpha$ resolution---the tilt-corrected images use 4x upsampling, which is discussed in \cite{yu_dose-efficient_2022, yu_dose-efficient_2024} and \cite{varnavides2025relaxing}.
The sample is MOF NU-1000 (\cite{li2025atomically}) with a thickness of 3 nm, and so is nearly a phase object and is much thinner than the depth of field.
The axial BF-STEM image (Figure \ref{fig_simulated_MOF}a) uses only about 4\% of the total incident dose and so is dominated by Poisson noise.
tcBF (Figure \ref{fig_simulated_MOF}b) shows more contrast for the larger features of the object, but from the diffractograms we observe the finest resolvable spacing is worse than for the axial BF image. 
acBF (Figure \ref{fig_simulated_MOF}c) correctly recovers the phase contrast information, including the contrast changes on each individual virtual image, and so is able to achieve atomic resolution of the heavy Zr atoms in the linker. 
Some information is transferred beyond the spatial frequency corresponding to 1$\alpha$, indicating an improvement in resolution over CTEM imaging for these conditions, but most fine features are lost to noise at this low dose condition. For comparison, experimental reconstructions of the same sample (but with greater thickness) and using an aberration-corrected instrument can be found in \cite{li2025atomically} using multislice ptychography, as well as \cite{zeltmann2025non} and \cite{ma_information_2025} using tilt-corrected methods. 

\subsubsection{Experimental reconstructions of twisted bilayer}

In practice, with modern aberration correctors, it is rare to have dominating higher-order aberrations. 
Figure \ref{fig_phason} shows reconstructions of a 2D material under typical imaging conditions, i.e., mostly defocus. 
For comparison, we also show the same region from the multislice ptychography reconstruction in \cite{zhang2025atom}, which achieves superresolution to clearly indicate the positions of each atom. 
While the tcBF reconstruction of this dataset (Figure~\ref{fig_phason}b,f) shows information transfer nearly out to the 2$\alpha$ limit, the oscillatory nature of the PCTF makes the image uninterpretable, with contrast that does not directly reflect the sample potential.
Post-summation sign correction (Figure~\ref{fig_phason}c,g) improves the interpretability by suppressing the long tail oscillations of the point spread function, but still does not access the anti-Friedel scattering present in the dataset and so misses or excessively damps certain spatial frequencies.
In CTEM, where the image is available live, it is possible to tune the defocus by hand to optimize the contrast and ensure that the spatial frequencies in a crystalline sample do not fall in the zeroes of the CTF.
As live video-rate tcBF is not yet available (to our knowledge, though processing power on modern computers is almost certainly sufficient to do so with an optimized program), it can be challenging to optimize conditions in a similar manner.
Because the PCTF of acBF (Figure~\ref{fig_phason}d,h) has no zeroes and is relatively insensitive to the precise defocus, it does not exhibit such focus-dependent contrast changes. 
The acBF image correctly resolves the fine spacings between slightly overlapped columns in the twisted bilayer, which is especially evident in the regions highlighted by the yellow boxes in Figure~\ref{fig_phason}.

With large apertures and residual aberrations, more often the limitation is poor probe coherence due to energy spread of the source.
Direct ptychography is known to be relatively robust to chromatic effects, with information transfer occurring strongly along the achromatic lines in the doubly reciprocal dataset \cite{pennycookHighDoseEfficiency2019}.
\cite{nellistConventionalInformationLimit1994} explored the various relevant envelope functions that incoherence produces in the 4D dataset, providing expressions for focal spread, finite detector size, and finite source size. 
The chromatic envelope they derive does not depend on the higher order aberrations or on the central defocus value, so acBF should be equivalently robust to this source of incoherence as its in-focus counterparts. 
When using an aberrated probe, the more rapid oscillation of the CTF produces greater sensitivity to detector size, requiring finer sampling of the diffraction plane. 
Related derivations for the tilted-beam CTEM case can be found in \cite{kosterAutotuningTEMUsing1989}. 
In iterative ptychography, it is now common to use multiple incoherent probe modes in the forward model when operating with a defocused probe to account for the partial coherence (\cite{chen2020mixed}). 

\subsubsection{Shift estimation}
When only defocus is present (and with a modern aberration corrector we should be able to ensure such conditions), the image shift from defocus has a simple visual interpretation as a parallax effect.
From a geometric optics standpoint, tcBF is equivalent to the shadow image montage method proposed by \cite{seiferShadowMontageConeBeam2025}.
However, the interpretation of the Ronchigram as a shadow image only holds directly when defocus is the only aberration; when other aberrations are present, the magnification of the shadow image is non-uniform across the bright-field disk and in general anisotropic (the local magnification is the Hessian of the aberration function evaluated at each detector position, $M(\bm\Theta)=\nabla^2\chi(\bm\Theta)$ (\cite{hawkesPrinciplesElectronOptics2018}, Eq. 41.25).
The distortion correction applied in acBF with higher order aberrations can thus also be interpreted as correcting for the nonuniformity of magnification in the Ronchigram prior to stitching the images.
The presence of these distortions also complicates the experimental determination of the shifts between virtual BF images, and thus the estimation of the aberration parameters. 
\cite{kosterAutotuningTEMUsing1989} considered the analogous situation in CTEM tilted-beam imaging, and found that ordinary cross-correlation of images yielded inaccurate shifts; their proposed solution instead fits the phase of the cross-correlation to a polynomial model, which is a similar approach as used in aberration-corrected single sideband ptychography (\cite{yang2016simultaneous}). 
In the tilt-corrected imaging methods, these challenges can likely be sidestepped by only cross-correlating the virtual BF images with those from nearby pixels on the detector, so that the distortions do not change much between the two images, combined with appropriate regularization of the shifts (\cite{yu_dose-efficient_2024, varnavides_iterative_2023}).
In \hyperref[appendix_C]{Appendix C} we explore the impact of imperfect shift estimation on the tilt-corrected imaging methods using simulations of the single atom. 
Choosing the shifts that maximize the image contrast produces complicated tails in the PSF.
It is common practice to fit an aberration function to the measured shifts and use the shifts given by the gradient of the best-fit aberration surface (\cite{varnavides_iterative_2023}); even when spherical aberration is present, satisfactory reconstructions can be produced by only considering the defocus term in the fit. 

\begin{figure*}[ht]
\centering
\includegraphics[width=0.55\linewidth]{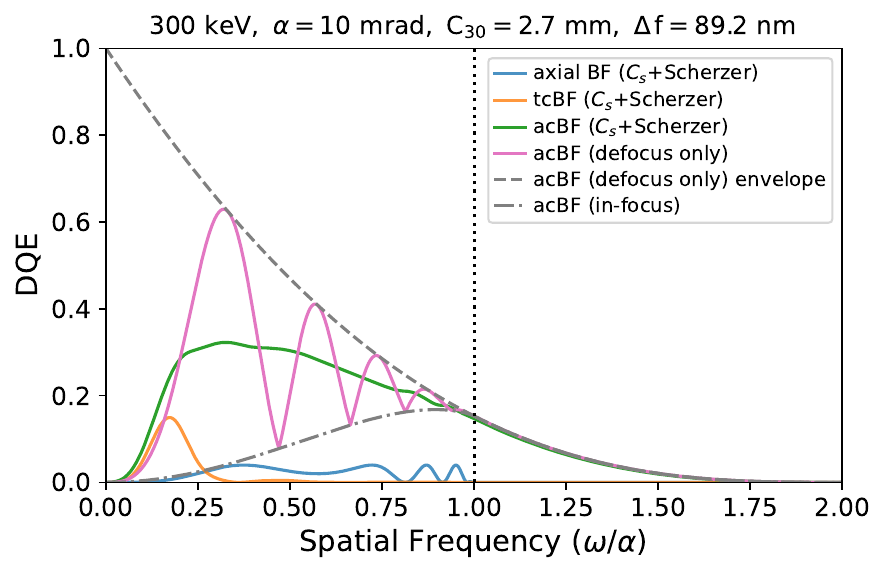}
\caption{Calculated detective quantum efficiency (DQE) of axial BF, tcBF and acBF for the same conditions as Figure~\Ref{fig_simulated_MOF} compared to acBF with and without $C_s$. The DQE of axial BF is low because it integrates over a small fraction of the bright-field disk. Similar to the PCTF in Figure \ref{1d_2d_ctf_sch_stig}, at low frequencies $[0,\alpha]$, the DQE of acBF with spherical aberration is smoothened and free of oscillations, roughly following the mean of the DQE of acBF with defocus only.}\label{1d_dqe_sch_stig}
\end{figure*}

\begin{figure*}[ht]%
\centering
\includegraphics[width=0.9\linewidth]{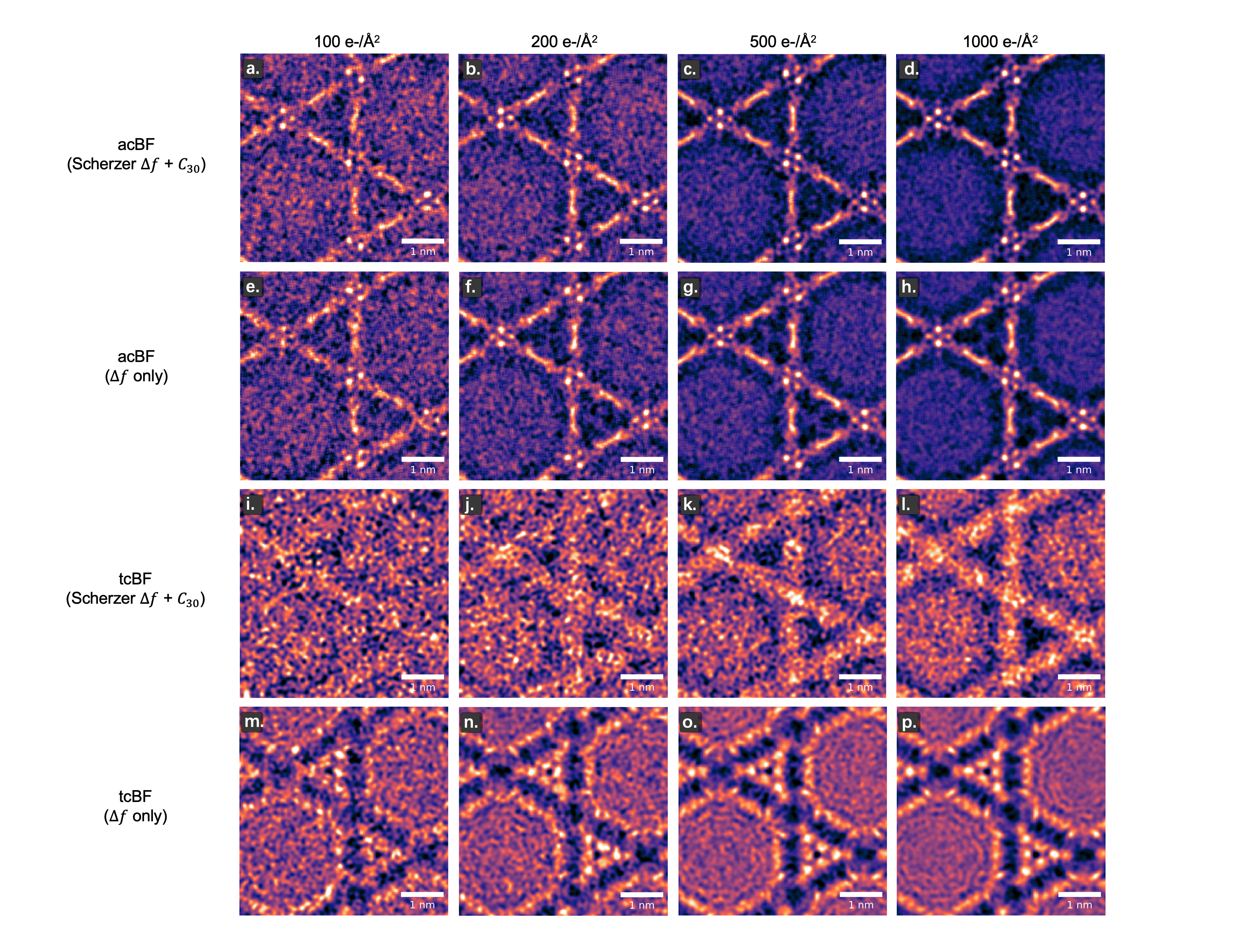}
\caption{AcBF imaging with combined spherical aberration and defocus (a-d) versus defocus only (e-h) of the same 3-nm-thick MOF sample as in Figure \ref{fig_simulated_MOF} at 300 keV and 10 mrad convergence semi-angle at different dose levels. AcBF with spherical aberration appears to produce generally more uniform contrast for the same information limit, with reduced ringing artifacts around the molecular chains. For comparison, tcBF imaging (i-p) is unable to resolve the Zr clusters and molecular chains in either case.}\label{acBF_dose}
\end{figure*}

\subsubsection{Improved information transfer at Scherzer conditions}

So far, we have compared PCTFs of various imaging modes and presented a simulated MOF imaging experiment under low dose condition. Here we extend the discussion to acBF with different aberrations and dose levels and examine the detective quantum efficiencies.
Figure \ref{1d_dqe_sch_stig} shows the DQE response for the different imaging modes with and without spherical aberration.  
We have accounted for the collection efficiency of the axial BF-STEM image, assuming a collection angle of 2~mrad. 
The tilt-corrected imaging methods all integrate over the full bright-field disk.
Defocused acBF with and without spherical aberration both produce continuously nonzero contrast transfer within $2\alpha$ well beyond the in-focus case, whereas the former is able to produce a more uniform contrast because of the smoother DQE curve. 
The tcBF with spherical aberration has poor information transfer and almost no usable signal beyond 0.25~$\alpha$. 

The importance of defocus can be seen by examining the in-focus responses which show a very reduced low-frequency response in Figure \ref{1d_dqe_sch_stig}. The DQE for in-focus acBF is approximately the same as for SSB, Wigner distribution deconvolution ptychography, and iCoM imaging, though slight differences in the noise propagation in each method causes them to vary from one another (\cite{bennemannDetectiveQuantumEfficiency2025}).

We further compare simulated acBF reconstructions with and without spherical aberration against tcBF in Figure~\ref{acBF_dose} for a range of doses. 
The acBF image contrast inside the MOF is similar for both aberration cases, with perhaps slightly clearer resolution of the Zr atoms at the vertices for the spherical aberration case compared to defocus only acBF, but the ringing artifacts around the molecular chains into the empty spaces are more noticeably reduced for the spherical abberation-optimized CTF as would be expected from its more uniform CTF. 
TcBF, on the contrary, is unable to resolve the Zr clusters or molecular chains in either case.

\subsubsection{Comparison with STEM phase plates}
The improvement in information transfer for the Scherzer condition can be understood by analogy to a STEM phase plate.
\cite{ophus2016efficient} investigated a STEM phase plate of the Fresnel type, with alternating zones of 0 and $\pi/2$ phase shift, where images are produced by summing the regions of the diffraction pattern with no shift and subtracting the shifted regions (MIDI-STEM).
In essence, this setup causes half the electrons to exactly meet the Zernike phase condition and half to have no phase contrast: the CTF climbs rapidly to about 0.5 at low frequencies and has a triangular falloff arising from the aperture overlap function.
\cite{hammel1995optimum} proposed a similar scheme where the objective lens aberrations produce the effective phase plate, though without the availability of 4D-STEM at that time the required detector setup was difficult to realize. 
Using a phase plate is more efficient since the exact phase condition can be met over half the detector area, while the phase plate produced by the lens aberrations can only approximate this condition, leading to some oscillations in the CTF.
\cite{yang2016enhanced} later combined the STEM Fresnel phase plate setup with ptychography reconstruction (ptychographic MIDI, or PMIDI), boosting the information transfer further by simultaneously accessing the anti-Friedel scattering component and thus improving the CTF in the mid- to high-spatial frequency ranges. 
It is notable that the CTF curves for PMIDI and Scherzer acBF are very similar.
Use of the Scherzer condition increases the range of scattering angles for which the correct phase condition is met for optimal interference contrast, producing a smooth CTF over a wider frequency range. 
Both methods approach the theoretical ideal information transfer for a phase plate in STEM, which is produced by random speckle phase modulation (\cite{dwyer_quantum_2024}).
The MIDI phase plate is operated in focus, which allows simultaneous ADF imaging but thus precludes faster acquisition by spatial upsampling of the reconstruction. 
For all STEM phase plates, accurate estimation of the phase modulation is necessary for reconstruction.
Physical phase plates can be difficult to characterize and will drift over time as material contamination increases, while using aberrations to induce phase modulations is more stable, does not require special hardware, and allows the phase shifts be determined self-consistently from the dataset via the tcBF shift measurement protocol.

\section{Conclusion}\label{Conclusion}

In this paper, we explored aberration-corrected bright-field STEM (acBF) as a maximally-efficient direct ptychography method for using information within the bright-field disk, by combining both the symmetric (tcBF) and antisymmetric (tcDPC) components of a 4D-STEM data set. 
Defocus is critical for accessing the symmetric component of the scattering, which provides most of the low-frequency information transfer; this information is missing in the direct ptychographies that are operated in focus, e.g. iDPC and SSB.
We analyzed the contrast transfer of off-axial bright-field STEM images in the presence of non-round and higher-order aberrations and find that the non-round aberrations produce anisotropic information transfer that cannot be recovered. 
Further, higher-order aberrations introduce complicated distortions and cross-terms between orders that vary for each off-axial image within the dataset.
Spherical aberration, which is prominent in cryo-electron microscopes, causes each off-axial image to have a different effective defocus, strongly damping the CTF of tcBF. 
By correcting the contrast transfer and distortions of each off-axial image before summation, acBF is able to treat these images as an effective focal series and recover a smoother CTF, which at Scherzer defocus approaches that of an ideal STEM phase plate. 
As a result, it yields interpretable phase-contrast images which maximally utilize the information encoded in the dataset under the weak phase approximation.

We demonstrate acBF reconstructions on several datasets to show its performance both at low dose and at high resolution and find that it outperforms tcBF in all cases.
Using simulated data from a MOF sample at low dose, we show that acBF is able to resolve the metal atoms in the nodes and show the structure of the organic linkers. 
Using an experimental dataset from a twisted bilayer of transition metal dichalcogenides, we show that acBF yields interpretable phase contrast images which correctly locate the atomic positions, while tcBF images alone do not correspond to the atomic structure as a result of its strong oscillations in the CTF from $\alpha$ to $2 \alpha$ - features that are removed in acBF. 
By self-consistently measuring and subsequently deconvolving the effect of probe aberrations, acBF acts as a straightforward, maximally efficient and computationally simple direct ptychography method for defocused 4D-STEM data.

\section{Data and code availability}
For review purposes, a minimal working example of the code and datasets are available at: \url{https://github.com/dsmagiya/acBF-STEM}.

\section{Competing interests}
No competing interest is declared.

\section{Acknowledgments}
DM is supported by the Center for Bright Beams, a U.S. National Science Foundation
Science and Technology center under Award PHY-1549132. SEZ and DAM are supported
by PARADIM under NSF Cooperative Agreement No. DMR-2039380. Electron microscopy
facility support is from the Cornell Center for Materials Research (CCMR) Shared Facilities and
PARADIM. CCMR facilities were supported through the NSF MRSEC program (DMR-1719875).
In addition, the authors want to thank Paul Cueva (NSF PHY-1549132) for his early work on
aberration and tilt measurements in 4D-STEM, Xavier
Baza for his tcBF simulations, and Guanxing Li for discussions on the MOF simulations. The authors acknowledge Colum O’Leary and Peter Nellist for
helpful discussions on direct ptychography; Georgios Varnavides and Colin
Ophus for helpful discussions on tcBF algorithms; and Peter Nellist, Angus Kirkland, Colin Ophus,
and Felix Bennemann on noise and information transfer. 
We also thank John Grazul, Mariena
Sylvestry Ramos and Phil Carubia for help with the instruments.

\bibliographystyle{apalike}
\bibliography{acBF}

\clearpage
\FloatBarrier  %

\clearpage
\onecolumn   
\renewcommand{\thefigure}{A\arabic{figure}}
\setcounter{figure}{0}
\renewcommand{\theequation}{A\arabic{equation}}
\setcounter{equation}{0}

\section{Appendix A. Tilt-correction for higher order aberrations}\label{appendix_A}

We start with the tilted-illumination CTFs in Equation \ref{tilt_pctf} and the wavefront aberration function defined by \cite{haider1998spherical}, expanded in a power series involving different orders of symmetries and multiplicities. 
To simplify the mathematical treatment of non-round aberrations, it is useful to use the complex convention for the aberration function. 
The Haider notation for the aberration coefficients are used here. 
A list of commonly used notations and the correspondence between them is available from \cite{sawada2008measurement}.
The aberration function in complex notation is written as
\begin{equation}\label{rose_chi}
\begin{aligned}
\chi(\omega)
= & \frac{2 \pi}{\lambda}  \operatorname{Re}  \left\{\frac{1}{2} \bar{\omega}^2 A_1+\frac{1}{2} \omega \bar{\omega} C_1+\frac{1}{3} \bar{\omega}^3 A_2+\omega^2 \bar{\omega} B_2\right. \\
& +\frac{1}{4} \bar{\omega}^4 A_3+\frac{1}{4}(\omega \bar{\omega})^2 C_3+\omega^3 \bar{\omega} S_3 \\
& +\frac{1}{5} \bar{\omega}^5 A_4+\omega^3 \bar{\omega}^2 B_4+\omega^4 \bar{\omega} D_4 \\
& +\frac{1}{6} \bar{\omega}^6 A_5+\frac{1}{6}(\omega \bar{\omega})^3 C_5+\cdots\}
\end{aligned}
\end{equation}
where the vectors for spatial frequency $\omega$ and tilt angle $\Theta$ are represented by complex numbers as
\begin{equation}
    \omega=|\omega| e^{i \phi}, \quad \Theta=|\Theta| e^{i \Phi}, \quad \Delta=\phi-\Phi,  
\end{equation}
\begin{equation}
    \omega=|\omega| \cos \phi+i |\omega|\sin \phi ,
\end{equation}
and the complex conjugate is
\begin{equation}
    \bar\omega = |\omega| \cos \phi - i |\omega| \sin \phi .
\end{equation}

Further, we note that the dot product of two vectors $\bm{a} \cdot \bm{b}$ becomes $\frac12\left( a \bar{b} + \bar{a} b \right)$ in the complex notation, and the squared magnitude $|\bm{\omega}|^2$ of a vector becomes $\omega \bar{\omega} = |\omega|^2$.
For the results in the main text, we use these identities to transform the equations back into the more familiar vector form. 

For brevity, we first write out the argument to the complex exponential in Eq.~\ref{tilt_pctf} using the representation \ref{rose_chi} for $\chi$. 
When Equation \ref{1d_ctf_df} is expanded and factored, the odd terms in the below expressions (those with $\pm$ in the below expressions) will factor out unconditionally (i.e. in both the TO and DO regions) and so produce the phase terms yielding image shifts and distortions.
The even terms (those that are the same for $+\Theta$ and $-\Theta$) do not factor out in both regions: they are the arguments to the $\sin$ part of the PCTF in the TO and give the phase modulation of the DO.

\subsection{Defocus ($C_1$)}
\begin{equation}\label{df_chi}
\begin{aligned}
\chi_{C_1}(\omega \pm \Theta)-\chi_{C_1}(\Theta)
&=\frac{2\pi}{\lambda} \frac{C_1}{2}[(\omega \bar{\omega}) \pm(\omega \bar{\Theta}+\bar{\omega} \Theta)] \\
 &=\frac{2\pi}{\lambda} \frac{C_1}{2}\left(|\omega|^2 \pm 2 |\omega| |\Theta| \cos \Delta\right) 
 .
\end{aligned}
\end{equation}
The first (odd) term in the brackets is the defocus part, and the second term is the image shift. 
Since there is no $\Theta$ dependence to the even part, the off-axial images have similar PCTFs and tcBF summation retains all symmetric contrast; the term resembles the original aberration function, so in this special case the tcBF PCTF has the same form as the axial PCTF.

\subsection{Spherical ($C_3$)}

\begin{equation}\label{C3_chi}
\begin{aligned}
\chi_{C_3}(\omega \pm \Theta)-\chi_{C_3}(\Theta)
&=\frac{2\pi}{\lambda}\,\frac{C_3}{4}\Big[(\omega \bar{\omega})^2
+(\omega \bar{\Theta}+\bar{\omega} \Theta)^2
+2(\omega \bar{\omega})(\Theta \bar{\Theta})
\;\pm\; 2(\omega \bar{\omega})(\omega \bar{\Theta}+\bar{\omega} \Theta)
\;\pm\; 2(\omega \bar{\Theta}+\bar{\omega} \Theta)(\Theta \bar{\Theta})\Big]\\[6pt]
&=\frac{2\pi}{\lambda}\,\frac{C_3}{4}\Big(|\omega|^4
+\big(4+2\cos 2\Delta\big)|\omega|^2 |\Theta|^2
\;\pm\; 4|\omega|^3|\Theta|\cos\Delta
\;\pm\; 4|\omega| |\Theta|^3 \cos\Delta\Big).
\end{aligned}
\end{equation}

The odd terms in the square brackets, are now notably different than the original aberration function: the $(\omega\bar\omega)^2$ term is the $C_3$, but additional terms which (after some manipulation) can be seen to have the form of defocus and two-fold astigmatism, appear. 
These give a complicated $\Theta$ dependence to the PCTF that tcBF does not account for but acBF does. 
The even terms also become more complicated, with a spatial-frequency dependent shift in addition to the expected simple spatial shift.

\subsection{Defocus + Spherical ($C_1$+$C_3$)}
In CTEM, it is common to achive optimal resolution by applying the Scherzer defocus to compensate $C_3$. 
Combination of any more than one aberration is straightforward with a linear summation, e.g. in this case adding up \ref{C3_chi} and \ref{df_chi}.
Doing so yields the PCTF given in the main text:
\begin{equation}
\begin{aligned}
\mathrm{PCTF}^{(C_1{+}C_3)}(\omega,\Theta)
&=\frac{i}{2\Omega_0}\,A(\Theta)\,
\exp\!\Bigg\{ i\,\frac{2\pi}{\lambda}\,|\omega||\Theta|\cos\Delta
\left[C_1+{C_3}\,(|\omega|^2+|\Theta|^2)\right] \Bigg\} \\
&\quad\times \Bigg[
A(\omega-\Theta)\,
\exp\!\Bigg(\!-i\,\frac{2\pi}{\lambda}\Big[
\frac{C_1}{2}|\omega|^2+\frac{C_3}{4}\big(|\omega|^4+(4+2\cos 2\Delta)|\omega|^2|\Theta|^2\big)
\Big]\Bigg) \\
&\qquad\;-\;
A(\omega+\Theta)\,
\exp\!\Bigg(\!+i\,\frac{2\pi}{\lambda}\Big[
\frac{C_1}{2}|\omega|^2+\frac{C_3}{4}\big(|\omega|^4+(4+2\cos 2\Delta)|\omega|^2|\Theta|^2\big)
\Big]\Bigg)
\Bigg].
\end{aligned}
\end{equation}

\subsection{Twofold astigmatism ($A_1$)}
\begin{equation}
\begin{aligned}
\chi_{A_1}(\omega \pm \Theta)-\chi_{A_1}(\Theta)
&=\frac{\pi}{\lambda}\,\frac12\Big(A_1\bar\omega^2+\bar A_1\omega^2\Big)
\ \pm\ \frac{\pi}{\lambda}\Big(A_1\bar\omega\bar\Theta+\bar A_1\omega\Theta\Big) \\[6pt]
&=\frac{\pi}{\lambda}\Big[
|A_1|\,|\omega|^2\cos\!\big(2(\phi-\psi_1)\big)
\ \pm\ 2|A_1|\,|\omega||\Theta|\cos(\phi+\Phi-2\psi_1)
\Big],
\end{aligned}
\end{equation}
where
\[
A_1=|A_1|\,e^{i2\psi_1},\qquad \psi_1\in[0,\pi).
\]
and $\psi_1$ sets the orientation of the twofold astigmatism axis.
Note that the odd part does not depend on $\Theta$, so each off-axial image has a similar PCTF to the axial image.
The even part contains only a shift term, which is the gradient of the aberration function as expected. 

\begin{equation}
\begin{aligned}
\mathrm{PCTF}^{(A_1)}(\omega,\Theta)
&=\frac{i}{2\Omega_0}A(\Theta)\,
e^{\,i\,\frac{\pi}{\lambda}\,2|A_1||\omega||\Theta|\cos(\phi+\Phi-2\psi_1)} \\
&\quad\times\Big[
A(\omega-\Theta)\,e^{-i\,\frac{\pi}{\lambda}\,|A_1|\,|\omega|^2\cos(2(\phi-\psi_1))}
- A(\omega+\Theta)\,e^{+i\,\frac{\pi}{\lambda}\,|A_1|\,|\omega|^2\cos(2(\phi-\psi_1))}
\Big].
\end{aligned}
\end{equation}

The second order aberrations produce dispersive shifts, and additionally produce an asymmetry between the two sidebands that complicates the information transfer. 
We note that these complications will greatly suppress tcBF contrast, while information can be largely---but, akin to the case of A1, not completely---recovered by acBF. 

\subsection{Threefold astigmatism ($A_2$)}
\begin{equation}
\begin{aligned}
\chi_{A_2}(\omega \pm \Theta)-\chi_{A_2}(\Theta)
&=\frac{\pi}{\lambda}\Big[
\tfrac13\big(A_2\bar\omega^3+\bar A_2\omega^3\big)
+\big(A_2\bar\omega\bar\Theta^2+\bar A_2\omega\Theta^2\big)
\ \pm\ \big(A_2\bar\omega^2\bar\Theta+\bar A_2\omega^2\Theta\big)
\Big] \\[6pt]
&=\frac{\pi}{\lambda}\Big[
\frac{2}{3}|A_2|\,|\omega|^3\cos\!\big(3(\phi-\psi_2)\big)
+2|A_2|\,|\omega||\Theta|^2\cos(\phi+2\Phi-3\psi_2) \\
&\qquad\qquad\quad
\ \pm\ 2|A_2|\,|\omega|^2|\Theta|\cos(2\phi+\Phi-3\psi_2)
\Big],
\end{aligned}
\end{equation}
where 
\[
A_2=|A_2|\,e^{i3\psi_2},\qquad \psi_2\in\Big[0,\tfrac{2\pi}{3}\Big).
\]
and $\psi_2$ sets the orientation of the threefold astigmatism.

\begin{equation}
\begin{aligned}
\mathrm{PCTF}^{(A_2)}(\omega,\Theta)
&=\frac{i}{2\Omega_0}A(\Theta)\,
e^{\,i\,\frac{\pi}{\lambda}\,2|A_2|\,|\omega|^2|\Theta|\cos(2\phi+\Phi-3\psi_2)} \\
&\quad\times\Bigg[
A(\omega-\Theta)\,e^{-i\,\frac{\pi}{\lambda}\left(\frac{2}{3}|A_2|\,|\omega|^3\cos(3(\phi-\psi_2))
+2|A_2|\,|\omega|\,|\Theta|^2\cos(\phi+2\Phi-3\psi_2)\right)} \\
&\qquad-\;
A(\omega+\Theta)\,e^{+i\,\frac{\pi}{\lambda}\left(\frac{2}{3}|A_2|\,|\omega|^3\cos(3(\phi-\psi_2))
+2|A_2|\,|\omega|\,|\Theta|^2\cos(\phi+2\Phi-3\psi_2)\right)}
\Bigg].
\end{aligned}
\end{equation}

\subsection{Coma ($B_2$)}
\begin{equation}
\begin{aligned}
\chi_{B_2}(\omega \pm \Theta)-\chi_{B_2}(\Theta)
&=\frac{\pi}{\lambda}\Big[
B_2(\omega^2\bar\omega+\Theta^2\bar\omega+2\omega\Theta\bar\Theta)
+\bar B_2(\bar\omega^2\omega+\bar\Theta^2\omega+2\bar\omega\bar\Theta\Theta) \\
&\qquad\qquad
\ \pm\ \big(B_2(\omega^2\bar\Theta+2\omega\Theta\bar\omega)
+\bar B_2(\bar\omega^2\Theta+2\bar\omega\bar\Theta\omega)\big)
\Big] \\[6pt]
&=\frac{\pi}{\lambda}\,2|B_2|\Big[
|\omega|^3\cos(\phi-\psi_B)
+|\omega||\Theta|^2\big(\cos(2\Phi-\phi-\psi_B)+2\cos(\phi-\psi_B)\big) \\
&\qquad\qquad
\ \pm\ \big(|\omega|^2|\Theta|\cos(2\phi-\Phi-\psi_B)
+2|\omega||\Theta|\cos(\Phi-\psi_B)\big)
\Big],
\end{aligned}
\end{equation}
where
\[
B_2=|B_2|\,e^{i\psi_B},\qquad \psi_B\in[0,2\pi).
\]
and $\psi_B$ sets the dipole axis of coma.

\begin{equation}
\begin{aligned}
\mathrm{PCTF}^{(B_2)}(\omega,\Theta)
&=\frac{i}{2\Omega_0}A(\Theta)\,
e^{\,i\,\frac{\pi}{\lambda}\,2|B_2|\big(|\omega|^2|\Theta|\cos(2\phi-\Phi-\psi_B)
+2|\omega||\Theta|\cos(\Phi-\psi_B)\big)} \\
&\quad\times\Bigg[
A(\omega-\Theta)\,e^{-i\,\frac{\pi}{\lambda}\,2|B_2|\Big(
|\omega|^3\cos(\phi-\psi_B)
+|\omega||\Theta|^2(\cos(2\Phi-\phi-\psi_B)+2\cos(\phi-\psi_B))\Big)}\,e^{\,i\,\phi_{B_2}} \\
&\qquad-\;
A(\omega+\Theta)\,e^{+i\,\frac{\pi}{\lambda}\,2|B_2|\Big(
|\omega|^3\cos(\phi-\psi_B)
+|\omega||\Theta|^2(\cos(2\Phi-\phi-\psi_B)+2\cos(\phi-\psi_B))\Big)}
\Bigg],
\end{aligned}
\end{equation}
with the extra constant phase (only for the ``$-$'' arm)
\[
\phi_{B_2}=\frac{4\pi}{\lambda}\,|B_2|\,|\Theta|^3\cos(\Phi-\psi_B).
\]

\newpage
\renewcommand{\thefigure}{B\arabic{figure}}
\setcounter{figure}{0}
\section{Appendix B. Tilt-tableaux of the PCTF}\label{appendix_B}
Here, we show detailed visualizations of the PCTFs for different aberration cases as tilt tableaux. 
Each subpanel of the figure is a 2D plot of $\operatorname{PCTF}(\bm\omega)$ at a given value of $\bm\Theta$, for tilts of 0, 15, and 30 mrad with various azimuths.
Note that this is a different projection of the 4D transfer function $\operatorname{PCTF}(\bm\omega,\bm\Theta)$ than is commonly shown in the SSB ptychography literature, where the plots are shown in $\bm\Theta$ space for a constant $\bm\omega$.
(In our plots the areas where the two circles overlap are the ``triple overlap'' and the areas with one circle are the ``double overlap.'' This nomenclature comes from the SSB convention; in the alternative projection, there are indeed three circles.)
\clearpage

\begin{figure}[H]%
\centering
\includegraphics[width=0.8\linewidth]{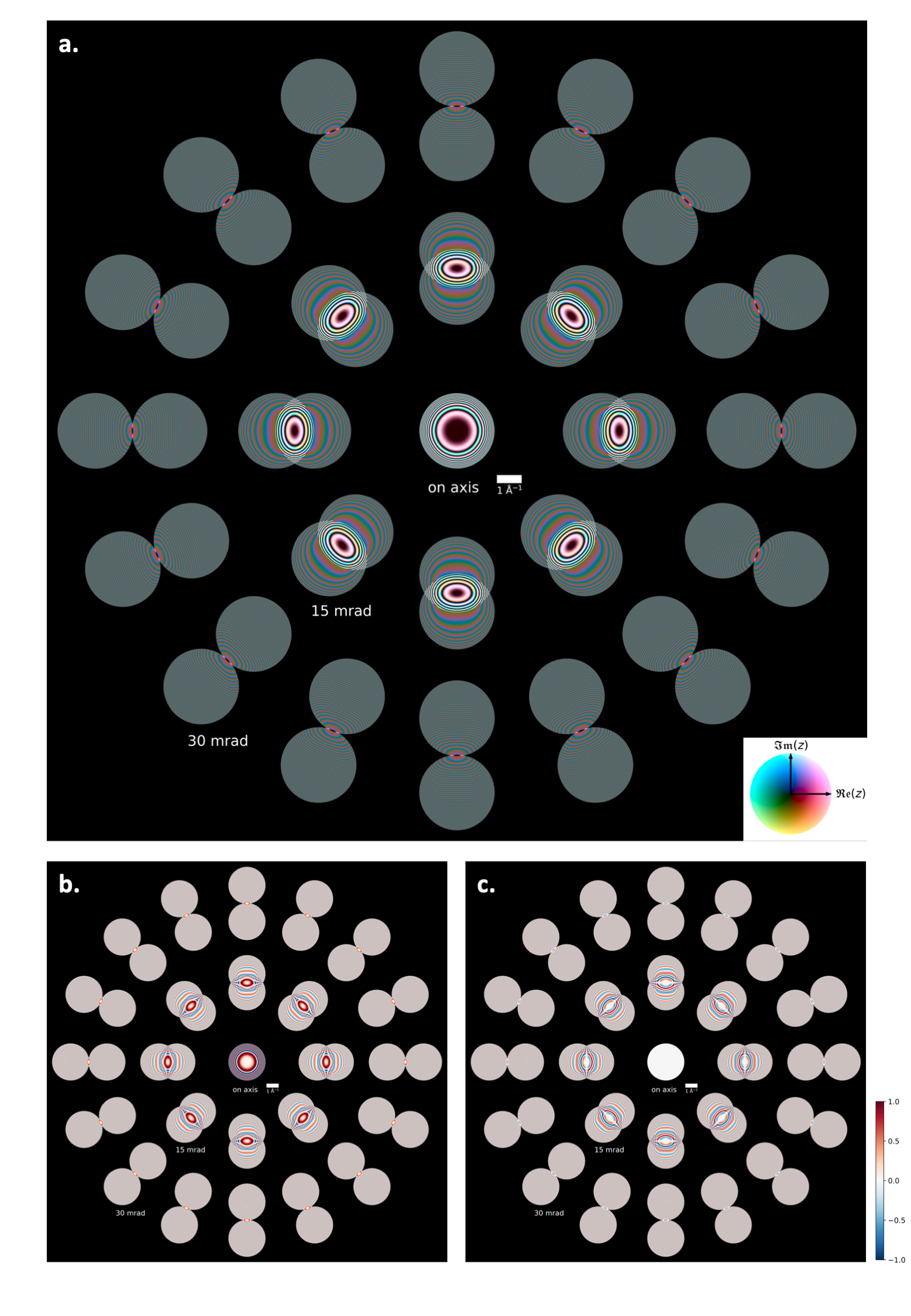}
\caption{Tilt-tableaux of the (a) complex, (b) real (symmetric) and (c) imaginary (anti-symmetric) components of the PCTF for 300 kV electrons with a 30 mrad semi-convergence angle and $C_3=50 \ \mu\text{m}$, extended to 30 mrad of tilt. }
\end{figure}

\begin{figure}[H]%
\centering
\includegraphics[width=0.8\linewidth]{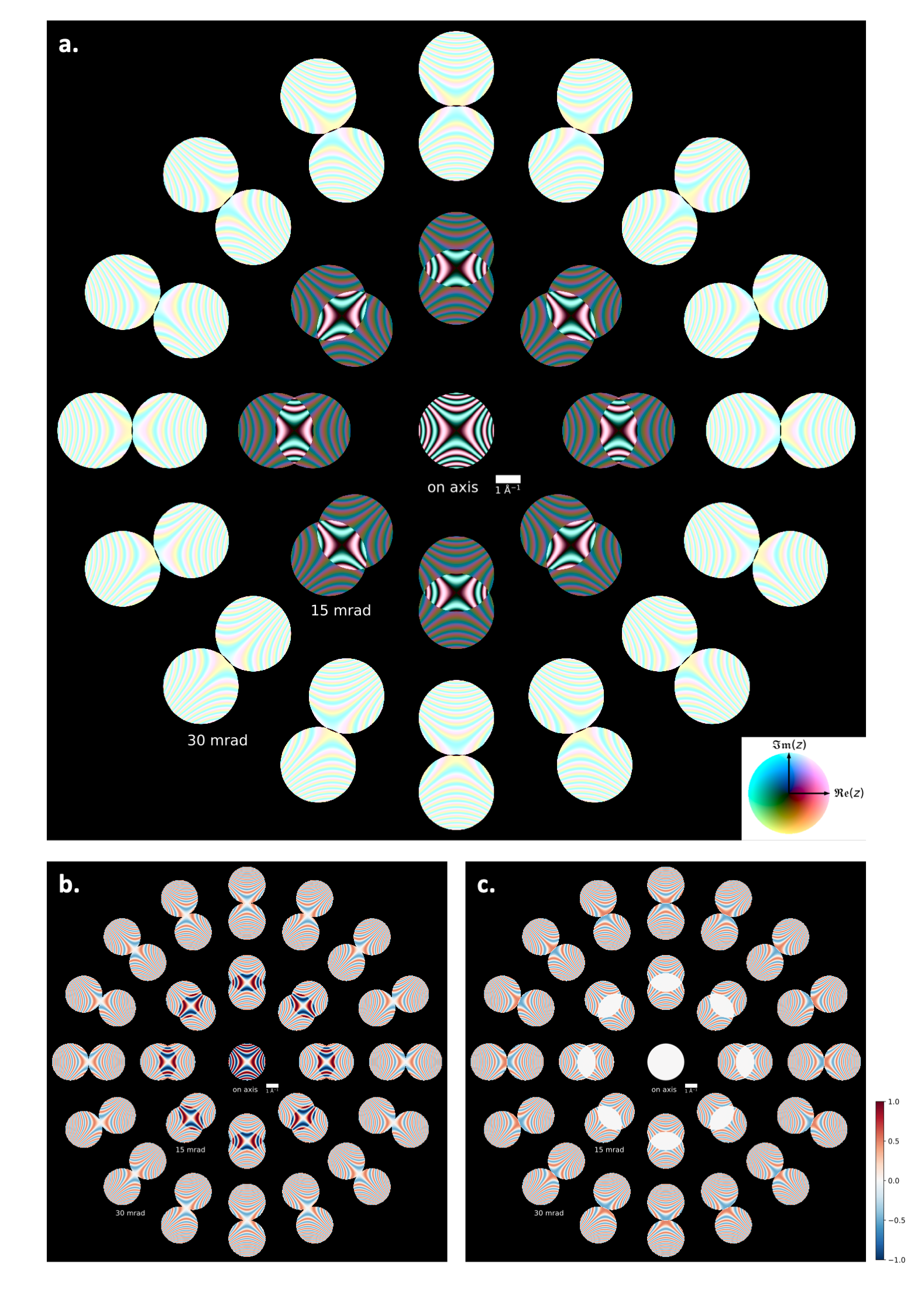}
\caption{Tilt-tableaux of the (a) complex, (b) real (symmetric) and (c) imaginary (anti-symmetric) components of the PCTF for 300 kV electrons with a 30 mrad semi-convergence angle and $A_1 =10 \ \text{nm}$, extended to 30 mrad of tilt. }
\end{figure}

\begin{figure}[H]%
\centering
\includegraphics[width=0.9\linewidth]{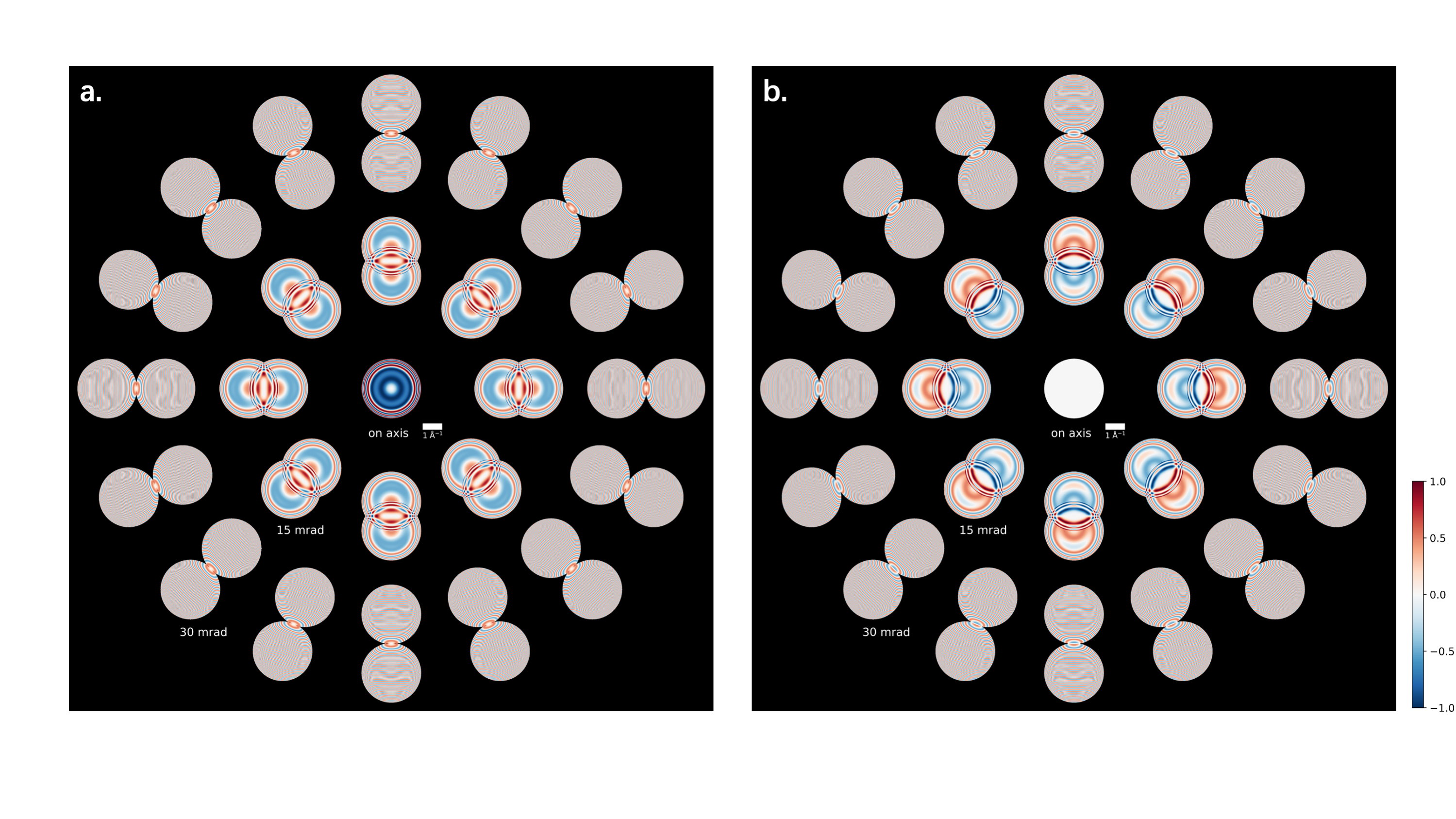}
\caption{Tilt-tableaux of the (a) real (symmetric) and (b) imaginary (anti-symmetric) components of the PCTF for 300 kV electrons with a 30 mrad semi-convergence angle, $C_3=50 \  \mu\text{m}$ and Scherzer defocus $\Delta f = 12.1 \ \text{nm}$, extended to 30 mrad of tilt. }
\end{figure}

\begin{figure}[H]%
\centering
\includegraphics[width=0.9\linewidth]{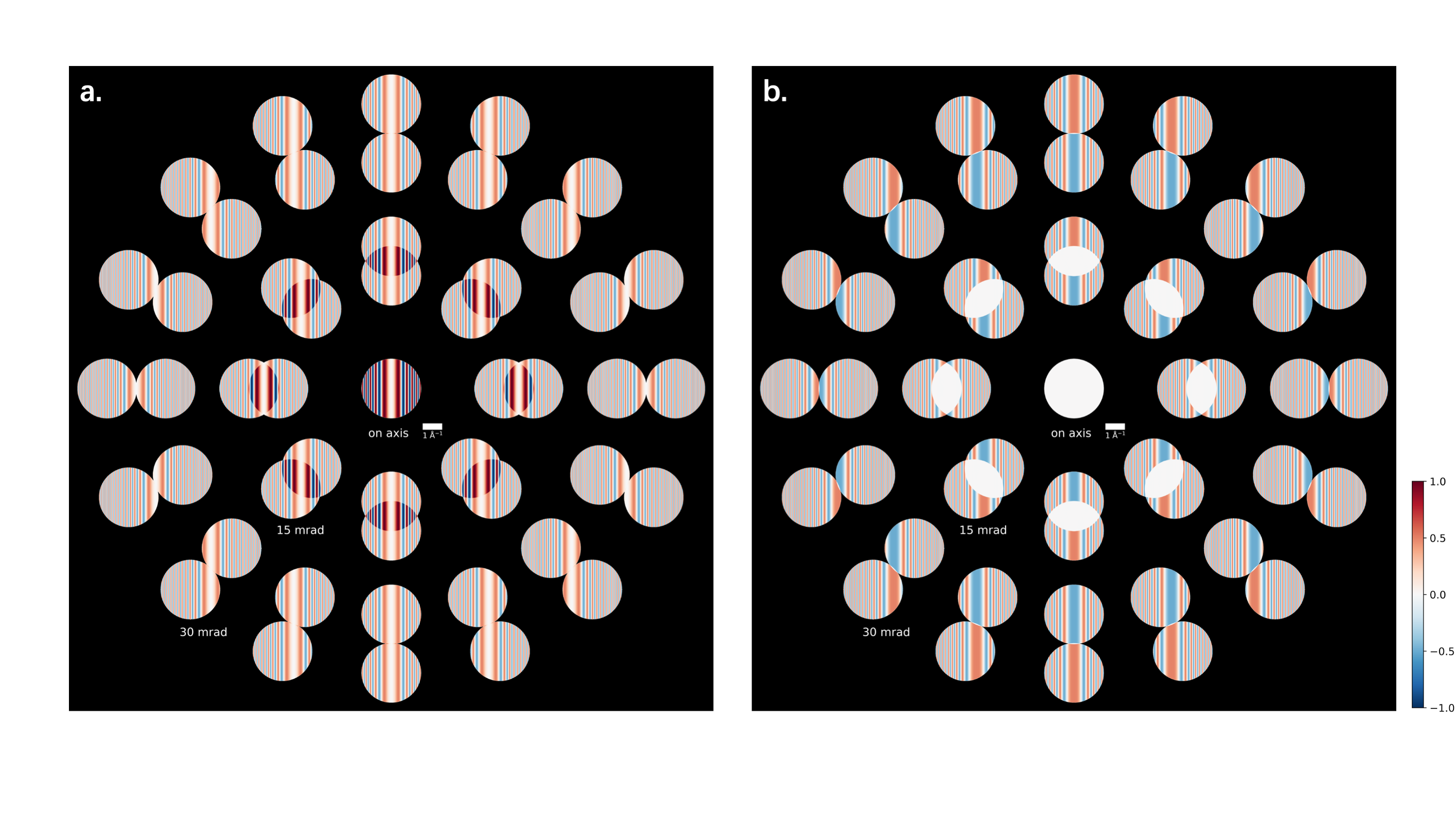}
\caption{Tilt-tableaux of the (a) real (symmetric) and (b) imaginary (anti-symmetric) components of the PCTF for 300 kV electrons with a 30 mrad semi-convergence angle, $A_1 =10 \ \text{nm}$ and defocus $\Delta f = -10\ \text{nm}$, extended to 30 mrad of tilt. }
\end{figure}

\renewcommand{\thefigure}{C\arabic{figure}}
\setcounter{figure}{0}
\section{Appendix C. Alternative shifting methods for tcBF}\label{appendix_C}

\begin{figure}[h]
\centering
\includegraphics[width=0.8\linewidth]{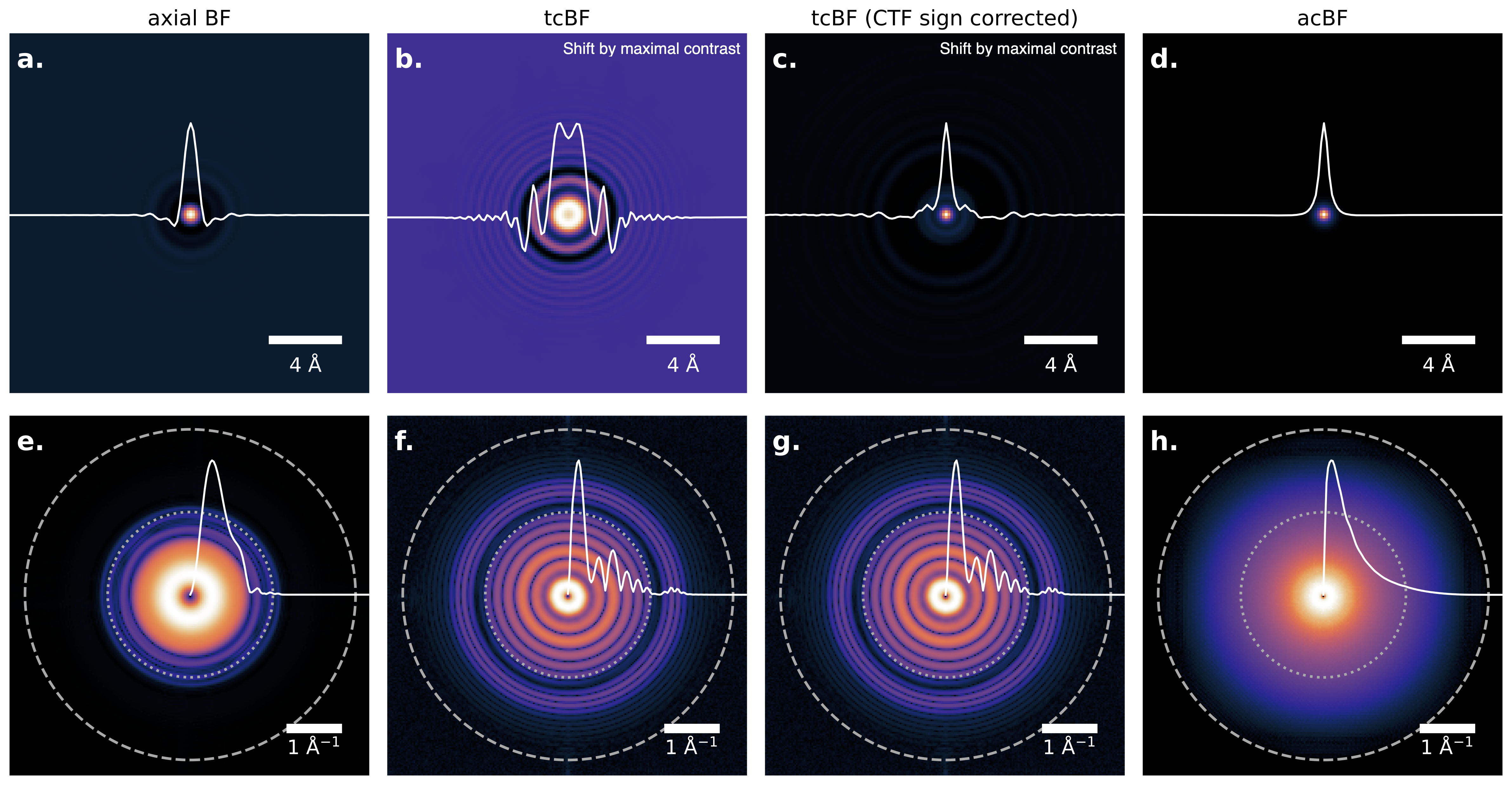}
\caption{Imaging of a simulated single C atom at infinite dose, 300 keV and 30 mrad convergence semi-angle
with $C_s$ = 50 $\mu$m and Scherzer defocus (121 Å) using different methods, (a) axial BF, (b) tcBF, (c) tcBF with post-summation CTF correction
(sign flipping), (d) acBF. (e–h) Corresponding fast Fourier transform (FFT) power spectra of (a–
d). The dashed circles on the FFTs indicate the spatial frequencies corresponding to 1$\alpha$ and 2$\alpha$. Instead of $\nabla \chi$, tcBF is shifted by local maximal contrast.}\label{fig_singleC_max}
\end{figure}

\begin{figure}[H]%
\centering
\includegraphics[width=0.8\linewidth]{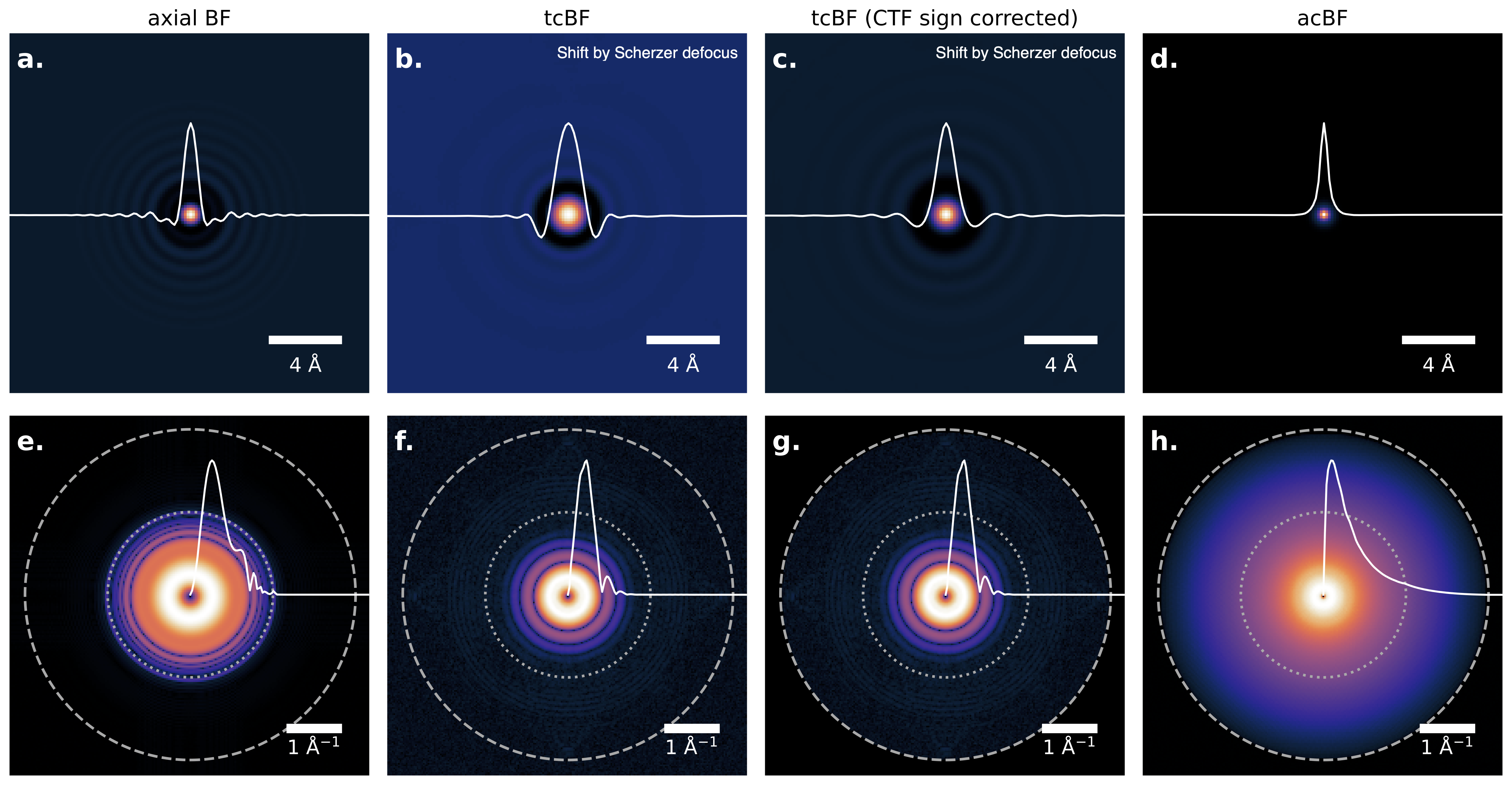}
\caption{Imaging of a simulated single C atom at infinite dose, 300 keV and 30 mrad convergence semi-angle
with $C_s$ = 50 $\mu$m and Scherzer defocus (121 Å) using different methods, (a) axial BF, (b) tcBF, (c) tcBF with post-summation CTF correction
(sign flipping), (d) acBF. (e–h) Corresponding fast Fourier transform (FFT) power spectra of (a–
d). The dashed circles on the FFTs indicate the spatial frequencies corresponding to 1$\alpha$ and 2$\alpha$. Instead of $\nabla \chi$, tcBF is shifted by the Scherzer defocus $\Delta f_\text{Sch}\bm\Theta$.}\label{fig_singleC_sch}
\end{figure}

In the main text, we have discussed the failure of tcBF when the off-axial images suffer from $C_{30}$-induced defocus changes. 
Here we show the CTF can be partly recovered by choosing shifts differently from those given by the gradient of the measured aberration function. 
For example, Figure \ref{fig_singleC_max} and \ref{fig_singleC_sch} show that the point resolution of tcBF can be improved by shifting by maximal local contrast and by shifting as if the only aberration was the defocus, respectively. 
These alternative strategies may be useful when it is difficult to measure the higher order aberrations correctly.

% \end{linenumbers}
\end{document}